\documentclass[a4paper,11pt]{article}
\usepackage{jheppub} % for details on the use of the package, please
\usepackage{amsmath,amssymb} 
\usepackage{subcaption}   

\usepackage{amsfonts}
\usepackage{enumerate}
\usepackage{ytableau}
\usepackage{blkarray}  
\usepackage{soul}
\usepackage{ulem}

\def\im{\text{i}}
\def\eqa{\begin{eqnarray}}
\def\eqae{\end{eqnarray}}
\def\be{\begin{equation}}
\def\ee{\end{equation}}
\def\bea{\begin{eqnarray}}
\def\eea{\end{eqnarray}}
\def\ba{\begin{array}}
\def\ea{\end{array}}
\def\bd{\begin{displaymath}}
\def\ed{\end{displaymath}}

\def\ie{{\it i.e.~}}

\def\>{\rangle}
\def\<{\langle}

\def\e{\epsilon}

\def\w{\omega}

\def\q{\theta}

\def\t{\tau}

\def\L{\Lambda}

\def\pa{\partial}

\def\bal#1\eal{\begin{align}#1\end{align}}

\def\({\left(}
	\def\){\right)}
	\def\[{\left[}
	\def\]{\right]}
	\def\nn{\nonumber \\}

\def\d{\operatorname{d}}
\def\Renyi{R$\acute{\text{e}}$nyi }
\def\Poincare{Poincar$\acute{\text{e}}$ }
\def\Banados{Ba$\tilde{\text{n}}$ados }

\title{\boldmath Spacetime Bananas with EOW Branes and Spins}

\author[a,b,1]{Jia Tian,}
\author[c]{Tengzhou Lai}
\author[b,2]{Farzad Omidi}

\affiliation[a]{State Key Laboratory of Quantum Optics and Quantum Optics Devices, \\ Institute of Theoretical Physics, Shanxi University, Taiyuan 030006, P.~R.~China}
\affiliation[b]{Kavli Institute for Theoretical Sciences (KITS),University of Chinese Academy of Science, \\ Zhongguancun East Road 80, Beijing 100190, P.~R.~China}
\affiliation[c]{School of Physical Sciences, University of Chinese Academy of Sciences, \\ Zhongguancun East Road 80, Beijing 100190, P.~R.~China}

\emailAdd{wukongjiaozi@ucas.ac.cn}
\emailAdd{laitengzhou20@mails.ucas.ac.cn}
\emailAdd{omidi@ucas.ac.cn}

\abstract{
In this work, we study and generalize the spacetime banana proposal for computing correlation functions of huge operators in the context of the  AdS$_3$/CFT$_2$ correspondence. First, we introduce time-like and space-like EOW branes into the proposal and demonstrate that: 1) a holographic dual of the one-point function in a BCFT can be obtained and its modified on-shell action reproduces the expected BCFT result; and 2) the GHY term on the stretched horizon can be replaced by the action of an EOW brane which wraps the horizon. Next, we discuss the two (one)-point function of huge spinning operators described by a rotating black hole in the bulk. We show that simply adding a GHY term on the stretched horizon is insufficient to reproduce the CFT results; instead, the appropriate modified action should be the micro-canonical action. Finally, we revisit the existing approaches for computing correlation functions using the gravity on-shell action of conical geometry or \Banados geometries. Surprisingly, we find that the on-shells actions of the \Banados geometries or the gravity solutions in the FG gauge yield unexpected incorrect results.
}

\arxivnumber{2410.18729}

\begin{document} 
	\maketitle
	\flushbottom

%\newpage

%\tableofcontents

%%%%%%%%%%%%%%%%%%%%%%%%%%
\section{Introduction}
%%%%%%%%%%%%%%%%%%%%%%%%%%
\label{section:introduction}
\renewcommand{\theequation}{1.\arabic{equation}}
\setcounter{equation}{0}
In the past three decades, various aspects of the AdS/CFT correspondence \cite{Maldacena:1997re,Witten:1998qj,Gubser:1998bc} have been explored. A crucial part of the dictionary is to understand how to reproduce the correlation functions, which are the fundamental observables in the CFT, from the information provided by the dual gravity theory. A celebrated statement regarding correlation functions involved with light (local) operators asserts that the generating functional of these correlation functions is equal to the partition function of dual string theory or the gravity theory in the saddle point approximation \cite{Witten:1998qj,Gubser:1998bc}:
\be 
\langle \exp\(\int d^d x\mathcal{O}\phi_{(0)}\)\rangle_{\text{CFT}}\approx e^{-I_{\text{bulk}}}\Big|_{\lim\limits_{z\rightarrow 0}\phi(z,x)z^{\Delta-d}=\phi_{0}(x)} \, ,
\ee 
where $\mathcal{O}$ is a light operator, $z\rightarrow 0$ denotes the asymptotic boundary of the AdS space, and $\Delta$ represents the conformal dimension of the operator. Given that the conformal dimensions of these light operators are very small, i.e. $\Delta \sim \mathcal{O}(1) \ll \frac{L_{AdS}^{d-1}}{G_N}$, their backreactions on the spacetime geometry are negligible. Consequently, the relevant bulk action simplifies to $I_{\text{bulk}}\approx S[\phi]$, and the correlation functions are computed using Witten diagrams \cite{Witten:1998qj,Freedman:1998tz}, which are the analogs of the Feynman diagrams in AdS space. Remarkably, in the $\mathcal{N}=4$ super Yang-Mills theory, the exact two-point functions of single-trace operators can be fully 
determined using integrability methods \cite{Beisert:2010jr}.
\\When the operator is heavy but not excessively so that its backreaction remains negligible (\ie its conformal dimension falls within the range $1 \ll \Delta \ll \frac{L_{AdS}^{d-1}}{G_N}$), the bulk action $S[\phi]$ can also be evaluated using a classical saddle point approximation. For instance, because the Green's function of scalar fields is equal to the path integral of a relativistic particle with the Polyakov action \cite{Cohen:1985sm} and in the classical limit the Polyakov action reduces to the length of the geodesic, the two-point function of heavy scalar operators can be calculated as follows \cite{Balasubramanian:1999zv,Louko:2000tp}:
\bea
\langle \mathcal{O}(x_1) \mathcal{O}(x_2) \rangle \sim e^{- m \ell },
\eea 
where $\ell$ is the regularized length of the geodesic inside the bulk spacetime, anchored at the insertion points of the operators on the boundary, and $m=\Delta$ is the AdS mass of the point particle. Similarly, considering that a heavy spinning operator is naturally dual to a spinning classical string, the two-point function of two heavy spinning operators \footnote{For a generalization to higher-points functions of heavy operators see \cite{Janik:2011bd,Buchbinder:2010vw,Buchbinder:2011jr,Zarembo:2010rr,Costa:2010rz,Klose:2011rm,Arnaudov:2011ek,Arnaudov:2011ivs,Minahan:2012fh,Bargheer:2013faa}.} can be computed using a classical approximation of the (modified) string  world-sheet Polyakov action \cite{Janik:2010gc} (See also Appendix \ref{appendix:string}). The corresponding classical string solution resembles a thickened geodesic or a ``world-sheet banana" connecting the two insertion points. A subtlety in this computation arises from the fact that the string has internal degrees of freedom, the Polyakov action has to be modified to accurately obtain a correlation function for operators with particular conformal dimensions.
\\Another interesting scenario occurs when the conformal dimensions $\Delta$ of the operators are of the order of $\frac{L_{AdS}^{d-1}}{G_N}$ 
but remain below the black hole threshold. The backreactions of such operators become significant, but only resulting in local alterations to the geometry of the bulk spacetime, such as conical singularities. Here, the correlation functions (or more precisely the conformal blocks) are captured by the carefully regularized on-shell actions of the backreacted gravitational solutions \cite{Chang:2016ftb,Chen:2016dfb}, although in practice constructing these backreacted solutions can be challenging \cite{Hadasz:2005gk,Harlow:2011ny,Faulkner:2013yia} \footnote{For a simple cut-and-glue construction of the backreacted AdS$_3$ geometry corresponding to two-point function, see \cite{Caputa:2022zsr}.}.
\\Inserting heavy operators with conformal dimensions exceeding the black hole threshold generates a black hole geometry. Recently, a novel holographic approach for computing the correlation functions of such operators, referred to as ``huge operators", was proposed in \cite{Abajian:2023jye,Abajian:2023bqv} (See also \cite{Kazakov:2024ald,Chandra:2024vhm}). It was demonstrated in \cite{Abajian:2023jye} that an AdS-Schwarzschild black hole can be transformed into a new geometry, termed ``spacetime banana", which describes a backreacted black hole connecting the two inserted operators. In particular, the black hole horizon itself adopts a banana shape. Furthermore, the on-shell action, after including a Gibbons-Hawking-York (GHY) term \cite{York:1972sj,Gibbons:1976ue} on the stretched horizon, is exactly equal to the two-point function (See eqs. \eqref{eq:action} and \eqref{two-point function-onshell action}). 
\\In this paper, we generalize the proposal for the banana geometry in asymptotically AdS$_3$ spacetimes from various perspectives, leaving the study to higher dimensions for future work. One advantage of working in 3D is the ease of comparing our results with the extensive literature on the AdS$_3$/CFT$_2$ correspondence. For example, the holographic \Renyi entropy, which is proportional to the logarithm of the two-point function of twist operators in a 2D CFT, is exactly equal \cite{Hung:2011nu} to the on-shell action of a specific \Banados geometry \cite{Banados:1998gg}.  It is noteworthy that the conformal dimension of the twist operator is already huge, $\Delta_n=\frac{c}{12}(n-\frac{1}{n})$, exceeding the BTZ black hole threshold $\frac{c}{12}$. Therefore, the \Banados geometry \eqref{twistmetric} appears to be a candidate geometry for computing the two-point function; however, it does not align with our intuitive understanding of a black hole connecting two insertion points, \ie a spacetime banana. To compare with the spacetime banana proposal, we revisit the calculation with the \Banados geometry proposal, we find that if we follow the standard approach in the literature, for example, as in \cite{Hung:2011nu} to compute the on-shell action $I_{\text{on-shell, FG}}$, defined in \eqref{EAdS}, for the \Banados geometry \eqref{twistmetric}, the partition function $e^{-I_{\text{on-shell, FG}}}$ does not equal the two-point function but rather its inverse. We further demonstrate that this result holds true for the conical geometry as well, namely, the on-shell action of the conical solution in Fefferman-Graham (FG) gauge is equal to the inverse of the two-point function. However, we emphasize that there is nothing inherently wrong with the \Banados geometry itself; the issue lies not with the geometry, but with the proposed action \eqref{EAdS}. Motivated by the spacetime banana proposal, we find that this issue can be resolved by including a missing portion of the spacetime in the FG gauge, as compared to the \Poincare AdS$_3$ spacetime.
\\Furthermore, it is interesting to note that without the GHY term at the stretched horizon,
the AdS-Einstein-Hilbert action of the spacetime banana is also equal to the inverse of the two-point function. This discrepancy arises from a reason similar to that in the string context: the black hole is dual to a canonical ensemble, not a specific state. Adding the GHY term plays the role of projecting the canonical ensemble onto a micro-canonical ensemble. Moreover, in the absence of the stretched horizon, the banana geometry is singular \cite{Abajian:2023jye,Abajian:2023bqv}; the stretched horizon thus acts as a physical boundary, and the GHY term ensures a well-defined variational principle. 
We find that the banana geometry can be regularized in another way by including an End-of-the-World (EOW) brane. Using an EOW brane to describe the micro-state of the black hole was first suggested in \cite{Kourkoulou:2017zaj,Almheiri:2018ijj} (also see \cite{Geng:2024jmm} for a recent discussion). Our proposal represents another realization of this idea. Besides this, we generalize the spacetime banana proposal by incorporating the EOW brane in the AdS/BCFT correspondence \cite{Karch:2000gx,Takayanagi:2011zk,Fujita:2011fp}, allowing us to compute the holographic correlation functions of huge operators in a BCFT.
\\In the original spacetime banana paper \cite{Abajian:2023jye}, it was also noted that the spacetime banana proposal can be generalized to other black holes with electric charge and matter. In this paper, we consider another natural generalization: a spinning spacetime banana. Analogous to the rotating string solution, the spinning spacetime banana should correspond to the correlation functions of two huge spinning operators. We find that the appropriate on-shell action for computing (part of) this correlation function is a refined version of the micro-canonical action \cite{Brown:1992bq}, which includes not only the GHY term at the stretched horizon but also an additional boundary term defined in \eqref{micbdy}. Notably, the new boundary term vanishes when the black hole angular momentum is zero, and hence, it is consistent with the non-rotating case. Since the micro-canonical action is universal and can be defined for all types of black holes, we anticipate that for spacetime bananas with other charges or matters, this modified action will remain suitable. However, using the current micro-canonical action, we are unable to reproduce the tensor structure associated with spin, as defined in \eqref{G2}. Note that the spinning banana geometry corresponds to a black hole whose angular momentum is parametrically as large as the central charge (see \eqref{eq:jbh}), leading to a highly non-trivial tensor structure that lacks an obvious gravitational realization. While 2D CFTs typically exhibit simpler tensor structures, in this case the result turns out to be complex, see \eqref{final_result}. Reproducing this structure from a real gravitational solution remains a challenge. Here, we propose a very preliminary attempt in this direction.
\\The outline of the paper is as follows: In Section \ref{section:banana} we first briefly review the original spacetime banana proposal,  focusing on the cone geometry, which is the middle step of the whole construction. We will make several comments and propose an alternative approach to modify the on-shell action. In Section \ref{section:EOW}, we explore the generalization of the spacetime banana proposal by incorporating the EOW branes. In Section \ref{section:spins}, we start from a rotating black hole and transform it into a spinning spacetime banana; we also review the concept of the micro-canonical action and propose a suitable refinement according to our setup, demonstrating that the micro-canonical action of the spinning spacetime banana indeed corresponds to the correlation function. We conclude the paper in Section \ref{section:discussion} with further comments and discussions. In Appendices \ref{appendix:twist} and \ref{appendix:particle}, we revisit the computations of on-shell action of \Banados geometries (AdS$_3$ solutions in FG gauge) and point out that there is a minus-sign discrepancy. The Appendix \ref{appendix:string} is dedicated to revisiting the string world-sheet banana proposal as a parallel comparison with the spinning spacetime banana result. Some technical computations about spacetime bananas with EOW branes are presented in Appendix \ref{appendix:bananaEOW}.

%%%%%%%%%%%%%%%%%%%%%%%%%%
\section{Spacetime Bananas}
%%%%%%%%%%%%%%%%%%%%%%%%%%
\label{section:banana}
\renewcommand{\theequation}{2.\arabic{equation}}
\setcounter{equation}{0}
In this section, we study the spacetime banana geometry introduced in \cite{Abajian:2023jye,Abajian:2023bqv}, which is holographically dual to the correlation function of two huge scalar operators.

\subsection{Review of the banana geometry}
%\label{section:banana}
The central idea of \cite{Abajian:2023jye,Abajian:2023bqv} is to foliate the \Poincare space in a novel manner, such that each foliation resembles a spacetime banana. This is accomplished through two consecutive coordinate transformations. We begin with the Euclidean non-rotating BTZ solution (or AdS-Schwarzschild black boles in higher dimensions):
\be \label{BTZ}
ds^2=(r^2-r_h^2)d\tau^2+\frac{dr^2}{(r^2-r_h^2)}+r^2d\phi^2,\quad r_h^2=8 G_NM-1,
\ee
where $G_N$ is the Newton constant and $M$ is the black hole mass. The first step is applying the Global-to-\Poincare (dubbed as GtP in \cite{Abajian:2023jye,Abajian:2023bqv}) map:
\bea \label{eq:goP}
e^{\tau+\im \phi}\frac{r}{\sqrt{1+r^2}}=T+\im X,\quad \frac{e^\tau}{\sqrt{1+r^2}}=Z.
\eea 
It should be pointed out that when $r_h^2=-1$, \eqref{BTZ} becomes the global AdS$_3$ metric and the map \eqref{eq:goP} will transform it to the \Poincare AdS$_3$ metric. On the other hand, for $r_h^2>0$, one obtains a deformed \Poincare AdS$_3$ spacetime with the metric given by 
\be  
ds^2=\frac{R^2 d\Theta^2}{Z^2}+\frac{(Z dR-R dZ)^2}{Z^2(R^2-r_h^2 Z^2)}+\frac{(R^2-r_h^2 Z^2)(RdR+Z dZ)^2}{Z^2(R^2+Z^2)^2},\label{eq:cone}
\ee
which asymptotically behaves as the \Poincare AdS$_3$ spacetime
\be 
ds^2 \sim \frac{dR^2+R^2d\Theta^2+dZ^2}{Z^2}+\mathcal{O}(Z^0),\quad \text{as }Z\rightarrow 0 .
\ee  
In this context, we introduce the polar coordinates {$(R, \Theta)$} defined by:
\bea 
&&T=R\cos\Theta,\quad X=R\sin\Theta,\\
&&r=\frac{R}{Z},\quad \tau=\frac{1}{2}\log(R^2+Z^2),\quad \phi=\Theta. \label{gtp2}
\eea 
Notably, \eqref{gtp2} maps the horizon $r=r_h$ in the BTZ coordinates to a cone defined by $R=r_h Z$ in the new coordinates, which we refer to as the cone coordinates. Therefore, the region inside of the cone ($R<r_h Z$) should be excluded. Moreover, the center of the black hole at the past infinity ($r=0,\tau=-\infty$) is mapped to the tip of the cone ($T=X=0,Z=0$). Therefore, the deformed \Poincare geometry can be interpreted as the back-reacted geometry with the insertion of a heavy local operator at the origin. To insert another heavy operator and derive the banana geometry, we apply a special conformal transformation (SCT)
\bea \label{stp}
&&T=\frac{t-b_t(x^2+t^2+z^2)}{1-2(b_x x+b_t t)+(b_x^2+b_t^2)(x^2+t^2+z^2)},\nn
&&X=\frac{x-b_x(x^2+t^2+z^2)}{1-2(b_x x+b_t t)+(b_x^2+b_t^2)(x^2+t^2+z^2)},\nn
&&Z=\frac{z}{1-2(b_x x+b_t t)+(b_x^2+b_t^2)(x^2+t^2+z^2)}.
\eea 
In these SCT coordinates $(t,x,z)$, the two insertion points are located at 
\be 
(t_1, x_1) = (0,0),\quad (t_2, x_2) = \( \frac{b_t}{b_x^2+b_t^2},\frac{b_x}{b_x^2+b_t^2} \)\equiv \vec{b},
\ee 
and the cone $(R=r_h Z)$ transforms into a banana defined by the two solutions $z_\pm$ of the following equation
\be 
r_h z=\sqrt{\left(t-{b_t} \left(t^2+x^2+z^2\right)\right)^2+\left(x-{b_x} \left(t^2+x^2+z^2\right)\right)^2}.
\ee 
The Euclidean gravity action considered in \cite{Abajian:2023jye} is
\be 
I_{\text{corr}}=\underbrace{-\frac{1}{16\pi G_N}\int d^3x\sqrt{g}(\mathcal{R}+2)}_{\text{AdS Einstein-Hilbert}}\underbrace{-\frac{1}{8\pi G_N}\int_{\partial_\text{AdS}}\sqrt{h}(\mathcal{K}-1)}_{I_{\text{GHY}}(\partial_{\text{AdS}})+I_{ct}}\underbrace{-\frac{1}{8\pi G_N}\int_{\partial_\text{stretch}}\sqrt{h_s}\mathcal{K}_s}_{I_{\text{GHY}}(\text{stretch})},\label{eq:action}
\ee 
where the stretched horizon $r=r_h+\epsilon_h$ is a boundary slightly outside the horizon, and the AdS cut-off surface is chosen to be $z=\epsilon$. It has been demonstrated in \cite{Abajian:2023jye} that the on-shell action of the banana geometry precisely reproduces the CFT two-point function:
\be 
e^{-I_{\text{corr}}}=\frac{1}{|\vec{b}|^{2\Delta}}=\langle \mathcal{O}(0)\mathcal{O}(\vec{b})\rangle,\quad \Delta=M.
\label{two-point function-onshell action}
\ee 
\subsection{One-point function}
The calculation of the on-shell action for the banana geometry is thoroughly presented in \cite{Abajian:2023jye}. Here, we provide additional details for the calculation for the on-shell action of the cone geometry, which is expected to correspond to the one-point function. In the next section, we will clarify this correspondence precisely within the AdS/BCFT framework. For the cone metric \eqref{eq:cone}, we find 
\be 
\sqrt{g}=\frac{R}{Z^3}, \label{determinant}
\ee 
which coincides with the determinant of  the \Poincare metric. We similarly choose the cut-off surface to be $Z=\epsilon$, with the normal vector given by $\vec{n}=-\vec{Z}$, leading to
\be 
\mathcal{K}=2+\mathcal{O}(\epsilon^4),\quad \sqrt{h}=\frac{R}{\epsilon^2}-\frac{(1+r_h^2)}{2R}.
\ee 
The stretched horizon is located at $R=(r_h+\epsilon_h)Z$, and the (unnormalized) normal vector $\vec{n}_s$ can be computed as follows
\be 
\vec{n}_s=- \left( \frac{d}{dZ},\frac{d}{dR},\frac{d}{d\Theta} \right) \frac{R}{Z}= \left(\frac{R}{Z^2},-\frac{1}{Z},0 \right). \label{eq:normalstretched}
\ee 
The resulting extrinsic curvature and the determinant of the induced metric at the stretched horizon are given by
\be 
\mathcal{K}_s=-\sqrt{\frac{r_h}{2\epsilon}},\quad 
\sqrt{h_s}=\frac{\sqrt{2r_h^3\epsilon}}{R},
\ee 
respectively. Substituting these ingredients into the action \eqref{eq:action}, we obtain
\bea  
I_{\text{corr}}&=&\underbrace{\frac{1}{4G_N}\int dR\(\frac{R}{\epsilon^2}-\frac{r_h^2}{R}\)}_{\text{AdS Einstein-Hilbert}}\underbrace{-\frac{1}{4G_N}\int dR\(\frac{R}{\epsilon^2}-\frac{(1+r_h^2)}{2R}\)}_{I_{\text{GHY}}(\partial_{\text{AdS}})+I_{ct}}\underbrace{+\frac{1}{4G_N}\int dR\frac{r_h^2}{R}}_{I_{\text{GHY}}(\text{stretch})}\nonumber\\
&=&\frac{(1+r_h^2)}{8G_N}\int_{\epsilon_r}^{\Lambda_r}\frac{dR}{R}=M\log \left( \frac{\Lambda_r}{\epsilon_r} \right), \label{trape}
\eea  
and thus,
\be 
e^{-I_{\text{corr}}}=\(\frac{\epsilon_r}{\Lambda_r}\)^{\Delta}\equiv \epsilon_r^\Delta\langle\mathcal{O}\rangle_{\text{disk}},\quad \Delta=M, \label{one}
\ee 
where we have introduced the UV regulator $\epsilon_r$ and IR regulator $\Lambda_r$. Due to the presence of the IR regulator $\Lambda_r$, which acts as a (cut-off) boundary, the partition function \eqref{one} exactly equals the one-point function on a disk with radius $\Lambda_r$.
As $\Lambda_r$ approaches $\infty$, the one-point function \eqref{one} vanishes, as expected due to the conformal symmetry. To obtain a non-trivial one-point function on a disk with a finite radius, we will cap the cone geometry with an EOW brane within the AdS/BCFT framework, as discussed in section \eqref{section:EOW}. Before proceeding, we would like to make a few remarks below.

\subsection*{Fefferman-Graham coordinates}
As discussed in Appendices \ref{appendix:twist} and \ref{appendix:particle}, the on-shell partition functions of the \Banados solutions are not equal to the correlation functions; rather, they are the inverses of these correlation functions. Let us examine the FG gauge of the cone geometry and banana geometry. Following \cite{Abajian:2023jye}, we find that the FG gauge of the banana geometry takes the same form as the 
\Banados solution \eqref{twistmetric}. Similarly, the FG gauge of the cone geometry is given by
\bea 
ds^2=\frac{dZ_f^2}{Z_f^2}+\frac{1}{Z_f^2}\(dU_f^2-\frac{(1+r_h^2)}{4\bar{U}_f^2}Z_f^2 d\bar{U}_f^2\)\(d\bar{U}_f^2-\frac{(1+r_h^2)}{4U_f^2}Z_f^2 dU_f^2\), \label{eq:FGcone}
\eea
which coincides with the \Banados solution \eqref{twistmetric} with \eqref{onetwist}, if we apply the following identifications 
\be 
Z_f=z,\quad U_f=u,\quad \bar{U}_f=\bar{u},\quad r_h^2=-\frac{1}{q^2}.
\ee 
As explained in Appendix \ref{appendix:twist}, the Einstein-Hilbert action (without the $I_{GHY}(\text{stretch})$) can be immediately computed by using \eqref{onfg}. For the cone geometry in the FG gauge, it is given by
\be 
I_{\text{cone-E-H,\,FG}}=-\frac{c}{6\pi}\frac{(1+r_h^2)}{4}2\pi \int_{\epsilon_r}^{\Lambda_r} \frac{dR}{R}=-\frac{(1+r_h^2)}{8G_N}\int_{\epsilon_r}^{\Lambda_r}\frac{dR}{R}=M\log \left( \frac{\Lambda_r}{\epsilon_r} \right) =-I_{\text{corr}}.
\ee 
Thus, we find ourselves in a familiar situation: the on-shell action has a sign problem or the partition function $e^{-I_{\text{cone-E-H,\,FG}}}$ actually equals the inverse of the correlation function. However, in this situation we can resolve this issue by adding the GHY term at the stretched horizon. A natural question arises: can we also resolve this issue in the previous two cases? An immediate obstacle is that there are no meaningful notions of the stretched horizon anymore. What should we add? 
\subsection*{Inside the banana}
In this section, we propose an alternative approach to cure the sign problem of the on-shell action. Our observation is that  $I_{\text{GHY}}(\text{stretch})$ is equal to the Einstein-Hilbert action of the region inside the cone or banana. Strictly speaking, the Euclidean geometry does not cover the region inside the horizon. In particular, the signature of the metric becomes $(-,-,+)$, obscuring the geometric interpretation of the spacetime within. However, the determinant of the metric \eqref{determinant} does not vanish anywhere, allowing us to extend the integral domain to the interior. Despite potential concerns about the singularity at the center of the black hole, it does not pose any technical difficulty, as we already introduce a cut-off disk around $R=0$ when evaluating the $R$ integral.
 As a quick check, the Einstein-Hilbert action of the extrapolated cone geometry is as follows
\be 
I_{\text{ex-cone}}=\underbrace{\frac{1}{4G_N}\int dR \( \frac{R}{\epsilon^2} \) }_{\text{AdS Einstein-Hilbert}}\underbrace{-\frac{1}{4G_N}\int dR\(\frac{R}{\epsilon^2}-\frac{(1+r_h^2)}{2R}\)}_{I_{\text{GHY}}(\partial_{\text{AdS}})+I_{ct}}=I_{\text{corr}}. \label{excone}
\ee 
This observation is significant because in all of the \Banados solutions (solutions in the FG gauge) we have discussed, the determinant of the metric vanishes at a certain surface, commonly referred to as a ``wall". When computing the on-shell action, we only include the spacetime region between the wall and the AdS boundary. Therefore, the FG gauge does not encompass the entirety of spacetime.
\\These two observations imply that to accurately reproduce the correlation function, we should also consider the regions that are excluded in the FG gauge. In Appendix \ref{appendix:particle}, we examine this proposal in the context of conical geometry. In global coordinates, the on-shell action should yield the conformal dimension of the heavy operator responsible for the conical singularity. By applying the GtP and SCT transformations, we can derive the analogs of the cone and banana coordinates, which enable us to compute the correlation functions correctly. We also rewrite these in the FG gauge and find that they only cover partial regions in global coordinates. For the analog of the cone geometry, we can identify the missing part of the spacetime in the FG gauge, which corresponds to a cone \eqref{misscone} in global coordinates. Consequently, the on-shell partition functions in the FG gauge do not match the correlation functions, and this discrepancy can be resolved by supplementing the on-shell partition function of the cone.
%%%%%%%%%%%%%%%%%%%%%%%%%%
\section{Spacetime Bananas with EOW Branes}
%%%%%%%%%%%%%%%%%%%%%%%%%%
\label{section:EOW}
\renewcommand{\theequation}{3.\arabic{equation}}
\setcounter{equation}{0}
In this section, we propose another possible way to regulate the spacetime banana by incorporating the EOW branes. We will make use of the EOW branes in two different ways. In the first scenario, akin to the standard AdS/BCFT formalism, the EOW brane intersects with the AdS boundary, thereby introducing a physical boundary in the dual CFT. This allows us to investigate the possibility that using the cone geometry and the banana geometry with EOW branes to compute the correlation functions in a BCFT.  In the second scenario, we consider a floating EOW brane (possibly with defects) in the bulk. We demonstrate that we can replace the $I_{\text{GHY}}(\text{stretch})$ with $I_{\text{EOW}}$ in \eqref{eq:action} to obtain the same result. 
\subsection{Review of the EOW brane}
An EOW brane is a codimension-one spacetime surface $\Sigma$ with the following action
\be 
I_{\text{EOW}}=-\frac{1}{8\pi G_N}\int_{\Sigma}\sqrt{h}(\mathcal{K}-T), \label{EOWaction}
\ee 
where $h_{ab}$ is the induced metric, $K_\Sigma$ is the extrinsic curvature and $T$ is the brane tension. The brane is dynamical and its profile is determined by 
\be 
\mathcal{K}_{ab}-h_{ab}(\mathcal{K}-T)=0, \label{eomeow}
\ee 
which amounts to imposing the Neumann boundary condition on the brane. We also allow the brane to have some sharp corners (defects) which can be realized by intersecting two smooth EOW branes \cite{Geng:2021iyq,Kusuki:2022ozk,Miyaji:2022dna,Biswas:2022xfw,Tian:2023vbi}. When the two branes have an intersection along the hypersurface $\Gamma$, we have to add the following term 
\be 
I_{H}=\frac{1}{8\pi G_N}\int_\Gamma(\theta-\theta_0)\sqrt{\gamma}, \label{Hayward}
\ee    
in the action which is the analog of the Hayward term \cite{Hayward:1993my} and $\theta_0$ is a fixed value that characterizes the corner. It should be pointed out that the inclusion of the defects is crucial for reproducing the BCFT spectrum \cite{Geng:2021iyq,Kusuki:2022ozk,Miyaji:2022dna,Biswas:2022xfw}. In the BTZ black hole metric \eqref{BTZ}, we can have two types of EOW branes, the "space-like`` one 
\be 
r(\phi)=\lambda \frac{r_h}{\sinh(r_h\phi)},\label{EOW:space}
\ee 
and the "time-like`` one
\be 
\sqrt{\frac{r^2}{r_h^2}-1}\cos(r_h\tau)=-\lambda, \label{EOW:time}
\ee
where $\lambda=-\frac{T}{\sqrt{1-T^2}}$.
\subsection{One-point function: time-like EOW brane}
\label{subsection:onepoint}
\begin{figure}[h]
\begin{center}
  \includegraphics[scale=0.5]{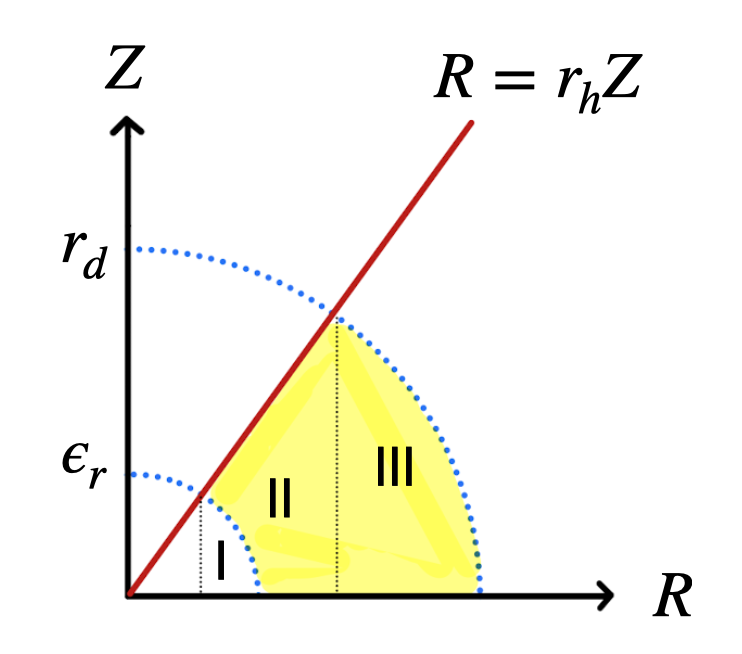}
  \caption{The holographic dual of the one-point function is the region shown in yellow bounded by the stretched horizon, AdS boundary, and two EOW branes. The red line denotes the horizon and the dotted blue lines denote the EOW branes. Moreover, the two vertical green lines are used to project the intersection points of the EOW brane and regulator surface with the horizon, to the boundary at $Z=0$.
  }\label{coneEOW}
  \end{center}
\end{figure}

The GtP transformation \eqref{gtp2} maps the time-like profile \eqref{EOW:time} to
\be 
\sqrt{1-\frac{ r_h^2 Z^2}{R^2}}\cos \left(\frac{1}{2}r_h\log[R^2+Z^2] \right)=-\lambda \frac{r_h Z}{R} \,,
\ee  
which intersects with the AdS boundary infinite times at 
\be 
R=e^{\frac{(2k+1)\pi}{2r_h}},\quad k\in \mathbb{Z}\, .
\ee 
To compute the one-point function, we can consider the tensionless case with $\lambda=0$, in which there are no additional contributions from boundary entropy since $I_{\text{EOW}}=0$. Moreover, the profile of the EOW brane simplifies significantly, taking the form of infinite hemispheres defined by
\be 
r_{\text{d}}=\sqrt{R^2+Z^2}=e^{\frac{(2k+1)\pi}{2r_h}},\quad k\in \mathbb{Z}\, .
\ee 
Among them we can choose any one to define the BCFT. Furthermore, we also include a tiny hemisphere with radius $\epsilon_r$ as the regulator of the origin of the disk. The configuration of the holographic dual geometry of the one-point function  is illustrated in Figure \ref{coneEOW}.
In this case, we divide the region outside the horizon which is enclosed by the EOW brane into four parts. Here we are interested in three regions denoted by I,II and III. It is obvious that the region in which the gravitational on-shell action reproduces the one-point function is II+III. In the trapezoidal region (I+II), the on-shell action can be computed exactly as shown in equation\eqref{trape}
\be 
I_{\text{I+II}} =  \frac{(1+r_h^2)}{8G_N}\int_{r_d \cos \alpha}^{\epsilon_r \cos \alpha} \frac{dR}{R}  =
M\log \left( \frac{r_{\d}\cos\alpha}{\epsilon_r\cos\alpha} \right) =M\log \left( \frac{r_{\text{d}}}{\epsilon_r} \right),
\ee  
where
\bea 
\alpha=\arctan \left( \frac{1}{r_h} \right),
\eea 
is the angle between the EOW brane and the boundary. It is obvious that the above on-shell action is already desirable to compute the one-point function. Therefore, it remains to demonstrate that the on-shell actions in regions I and III cancel each other. Indeed, we find that this is the case:
\bea  
I_{\text{III}}-I_{\text{I}}&=&\frac{(1+r_h^2)}{8G_N}\(\int_{r_{\text{d}}\cos\alpha}^{r_{\text{d}}}-\int_{\epsilon_r\cos\alpha}^{\epsilon_r}\)\frac{dR}{R} \nonumber \\
&+&\frac{1}{8G_N}\(\int_{\epsilon_r\cos\alpha}^{\epsilon_r}\frac{dR^2}{(\epsilon_r^2-R^2)}-\int_{r_{\text{d}}\cos\alpha}^{r_{\text{d}}} \frac{dR^2}{(r_{\text{d}}^2-R^2)}\)=0\, .
\eea   
One can also consider EOW branes with non-trivial tension, but in this case the on-shell action will also receive another contribution due to the boundary entropy.
\subsection{Towards two-point functions}
\begin{figure}[h]
\begin{center}
  \includegraphics[scale=0.5]{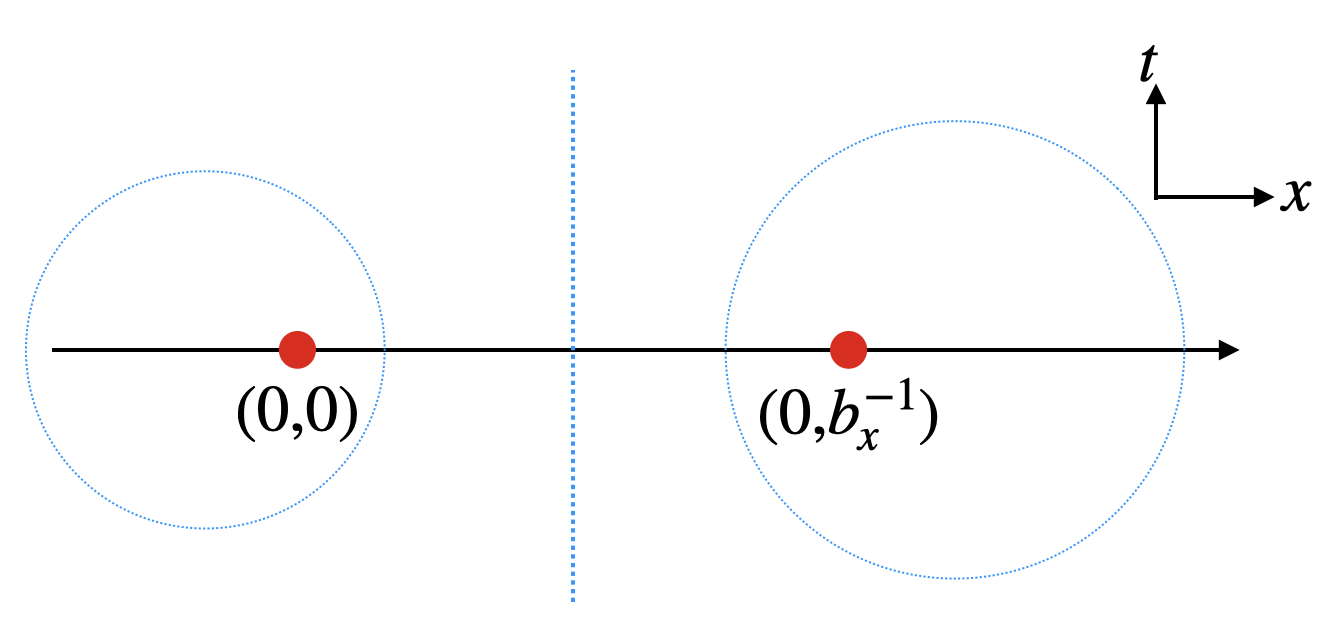}
  \caption{The EOW brane can only enclose one of the two inserted operators. Here the dashed blue circles are the projection of the EOW branes on the boundary. With the special value $b_x r_{\d }=1$, the EOW brane becomes a plane that cuts the banana in half. In this case, the projection of the  EOW brane on the boundary is shown by a vertical blue dashed line.}\label{bananaEOWpic}
  \end{center}
\end{figure}

The next step is to consider higher point correlation functions and finally establish a general formalism for computing correlation functions in BCFTs. Because the EOW brane has to satisfy the equation of motion \eqref{eomeow}, its profile is highly constrained. For example, in the \Poincare AdS$_3$ spacetime, the profile can only take the form of a plane or sphere. Although more complicated profiles can be achieved by introducing defects on the brane \cite{Kanda:2023zse,Liu:2024oxg}, such modifications also alter the CFT state, rendering them unsuitable for computing vacuum correlation functions.
\\In the following discussion, we will demonstrate that the SCT transformation described in \eqref{stp} does not yield a bulk solution for computing the two-point function on a disk. The SCT transformation maps a circle at the AdS boundary in the cone coordinates to another circle in the banana coordinates. Without loss of generality, we can set $b_t=0$, then the circle in the banana coordinates is described by the following equation
\footnote{This equation is valid for $b_x r_{\d}\neq 1$.}
\be 
\(x-\frac{b_x r_{\d}^2}{b_x^2r_{\d}^2-1}\)^2+t^2=\(\frac{r_{\d}}{b_x^2r_{\d}^2-1}\)^2. \label{circle}
\ee 
To compute the connected two-point function on a disk, it is crucial that the two insertion points $(0,0)$ and $(0,\frac{1}{b_x})$ stay within the same circle. This requirement leads to two conditions 
\bea  
\(\frac{b_x r_{\d}^2}{b_x^2r_{\d}^2-1}\)^2<\(\frac{r_{\d}}{b_x^2r_{\d}^2-1}\)^2,\quad &\rightarrow &\quad b_xr_{\d}<1 ,\\
\(\frac{1}{b_x}-\frac{b_x r_{\d}^2}{b_x^2r_{\d}^2-1}\)^2<\(\frac{r_{\d}}{b_x^2r_{\d}^2-1}\)^2,\quad &\rightarrow&\quad \frac{1}{b_xr_{\d}}<1.
\eea  
However, these two conditions conflict with each other. As a result, the solution from the SCT transformation can only be used to calculate the one-point function on a disk, such that the inserted operator is not necessarily  at the center. If we choose $b_x r_{\d }=1$, the circle becomes a line that runs through the midpoint of the two insertion points. Figure \ref{bananaEOWpic} illustrates the possible configurations of the EOW branes in the banana coordinates.
It suggests that this configuration corresponds to the one-point function in the half-plane. Since this planar EOW brane precisely cuts the banana in half, the on-shell action is also half of $I_{\text{banana}}$ in \eqref{eq:action}. In other words, we have
\be 
e^{-I_{\text{half banana}}}=\frac{1}{(1/b_x)^{\Delta}}=\langle \mathcal{O}(1/b_x)\rangle_{\text{UP}}.
\ee

\subsection{An alternative proposal: space-like EOW brane}
\begin{figure}[h]
\begin{center}
  \includegraphics[scale=0.5]{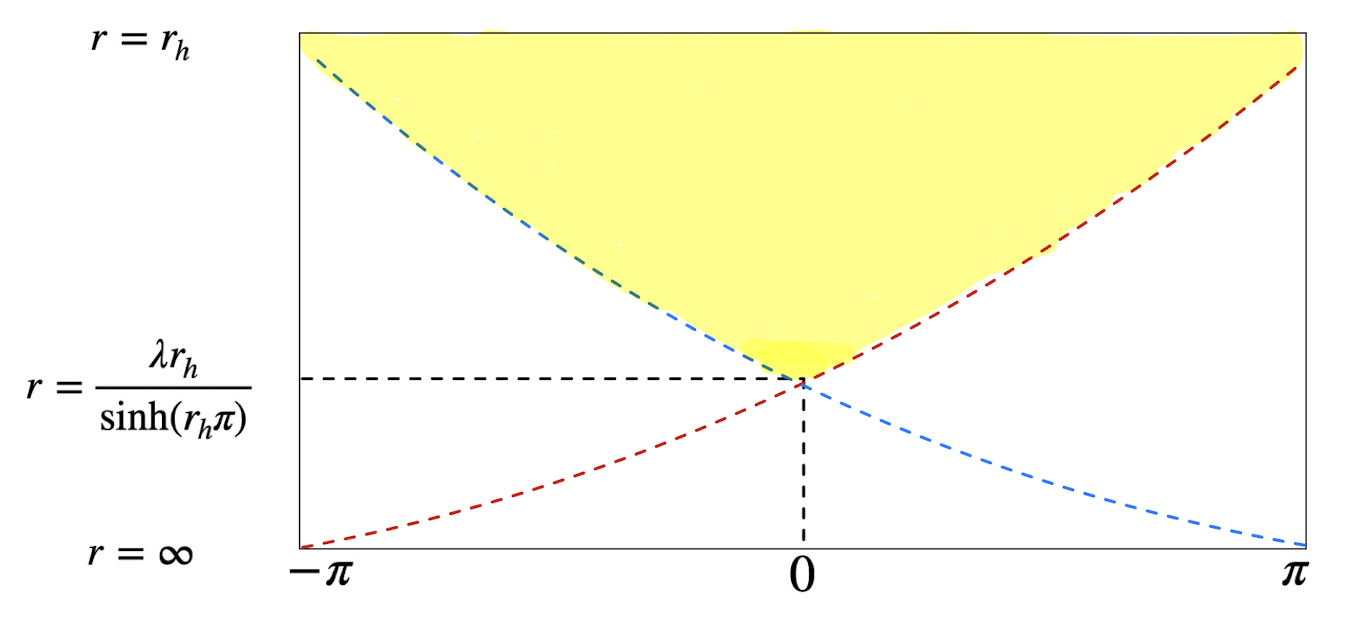}
  \caption{We can use two EOW branes to enclose the horizon of the black hole. These branes are shown by blue and red dashed curves. Moreover, the yellow region is outside the horizon and enclosed by the two EOW branes. Here we calculate the bulk on-shell action $I_{\rm EH}$ inside this region.}\label{eow1}
  \end{center}
\end{figure}

\begin{figure}[h]
\begin{center}
  \includegraphics[scale=0.9]{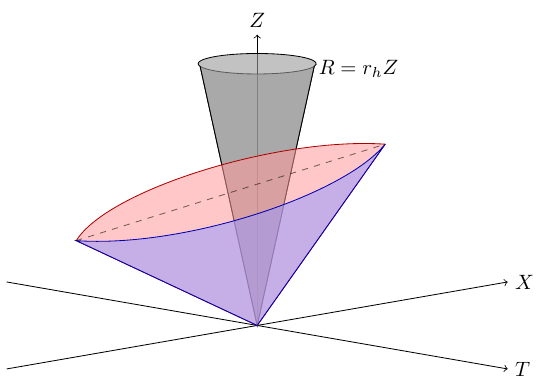}
  \caption{The two EOW branes form a tilted cone outside the cone corresponding to the stretched horizon.}\label{tcone}
  \end{center}
\end{figure}

In the spacetime banana proposal \cite{Abajian:2023jye}, the key points are the introduction of the stretched horizon and the addition of $I_{\text{GHY}}(\text{stretch})$ to the action. This term plays a crucial role in projecting the canonical ensemble, which is dual to the BTZ black hole, onto a specific micro-canonical state. In this subsection, we propose an alternative method for achieving this projection by placing an EOW brane in the bulk that encloses the horizon. The motivation for considering this configuration is twofold: First, in the original spacetime banana proposal, adding $I_{\text{GHY(stretch)}}$ corresponding to imposing Dirichlet boundary conditions at the stretched horizon.  However, this is not the only choice that yields a well-defined variational principle. We demonstrate that Neumann boundary conditions provide an alternative, equally consistent option. Second, following suggestions in \cite{Kourkoulou:2017zaj, Almheiri:2018ijj} that black hole microstates can be constructed by capping the geometry with an EOW brane, our construction here can be viewed as a realization of this proposal. 
\\The holographic model featuring EOW branes in the bulk has been explored in \cite{Akal:2020wfl}, where it is used to describe phenomena such as bubble nucleation in the universe or the nucleation of a bubble of nothing. However, our construction is more closely aligned with the approach in \cite{Wei:2024zez}, where the bulk EOW brane is introduced to regularize the orbifold singularity in the gravitational dual of the cross-cap state. 
\\We begin with the BTZ solution \eqref{BTZ} and demonstrate that the EOW brane action \eqref{EOWaction} is sufficient to realize the following transformation
\be 
Z=\text{tr}(e^{-\beta H})=e^{-\beta F_{\text{Gibbs}}}\rightarrow e^{-\beta E}. \label{freetr}
\ee 
Notice that the space-like EOW brane \eqref{EOW:space} extends from the AdS boundary to the interior of  the black hole. Therefore, to enclose a bulk subregion, we need at least two such EOW branes. For simplicity, we consider a symmetric configuration shown in Figure \eqref{eow1} which consists of two EOW branes with the following profiles
\be \label{tEOW}
r(\phi)=\begin{cases}
	\frac{\lambda r_h}{\sinh(r_h(\pi-\phi))}, & \phi\in[-\pi,0]\\
	\frac{\lambda r_h}{\sinh(r_h(\pi+\phi))}.& \phi\in[0,\pi]
\end{cases}
\ee   
These branes intersect at two points 
\be 
(r_1, \phi_1) = \(\frac{\lambda r_h}{\sinh(r_h\pi)},0\),\quad (r_2, \phi_2) = \(\frac{\lambda r_h}{\sinh(2r_h\pi)},\pi\).
\ee 
\\It should be pointed out that the intersection angles are not important to us, since the intersection term \eqref{Hayward} vanishes on-shell \cite{Miyaji:2022dna,Biswas:2022xfw,Tian:2023vbi}. To ensure that the EOW branes in \eqref{tEOW} remain outside the horizon, the brane tension must satisfy the following condition\footnote{The normal vectors of the EOW branes point inward, so the tension of the branes are negative.} 
\be 
-\frac{T}{\sqrt{1-T^2}} > \sinh(2\pi r_h),\quad \rightarrow \quad -1<T<-\tanh(2\pi r_h).
\ee 
The on-shell action of this kind of region has been computed in \cite{Tian:2023vbi}, yielding the following result  
\be 
I_{\text{BTZ+EOW}}=I_{\text{EH}}+I_{\text{GHY}}(\partial_{AdS})+I_{ct}+I_{\text{EOW}}+I_{H}=\beta \( M-\frac{c}{12} \)\equiv \beta E,
\ee 
as anticipated. 
Now we are ready to apply the GtP transformation \eqref{gtp2} which maps both of the EOW branes \eqref{tEOW} to a tilted cone:
\be \label{tEOWc}
Z=\begin{cases}
	R \( \frac{\sinh(r_h(\pi-\Theta))}{\lambda r_h} \), & \Theta\in[-\pi,0]\\
	R \( \frac{\sinh(r_h(\pi+\Theta))}{\lambda r_h} \),& \Theta\in[0,\pi]
\end{cases}
\ee
which is illustrated in Figure \ref{tcone}.
Let us focus on the on-shell action in the half cone which is described by $\Theta\in[0,\pi]$. The standard AdS gravity part is 
\bea  
I_{\text{AdS, half}}&=&\underbrace{\frac{1}{8G_N}\int_{\epsilon_r}^{\Lambda_r} dR\(\frac{R}{\epsilon^2}-\frac{R}{Z_*^2}\)}_{\text{AdS Einstein-Hilbert}}\underbrace{-\frac{1}{8G_N}\int_{\epsilon_r}^{\Lambda_r} dR\(\frac{R}{\epsilon^2}-\frac{(1+r_h^2)}{2R}\)}_{I_{\text{GHY}}(\partial_{\text{AdS}})+I_{ct}} \\
&=& \( \frac{M}{2} \) \log \( \frac{\Lambda_r}{\epsilon_r} \)-\frac{1}{8G_N}\int_{\epsilon_r}^{\Lambda_r} \frac{RdR}{Z_*^2},\label{eowbulkcone}
\eea   
where we defined $Z_*= \frac{R \sinh(r_h(\pi+\Theta))}{\lambda r_h}$. Given the EOW brane profile \eqref{tEOWc}, it is straightforward to obtain the induced metric and the extrinsic curvature. We find
\be 
K-T=\frac{1}{2}K=-\frac{\lambda}{\sqrt{1+\lambda^2}},\quad \sqrt{h}=\frac{r_h^2\lambda\sqrt{1+\lambda^2}}{R\sinh^2(r_h(\pi+\Theta))},
\ee 
which lead to the following EOW brane action
\be 
I_{\text{EOW}}=\frac{1}{8G_N}\int_{\epsilon_r}^{\Lambda_r} dR\frac{\lambda^2 r_h^2}{R\sinh^2(r_h(\pi+\Theta))}=\frac{1}{8G_N}\int_{\epsilon_r}^{\Lambda_r} dR \frac{RdR}{Z_*^2}.
\ee 
Therefore, the EOW brane action cancels the second term in \eqref{eowbulkcone}. After taking account of the contribution of the other half of the cone, we end up with
\be 
I_{\text{AdS}}=M\log \( \frac{\Lambda_r}{\epsilon_r} \),
\ee 
which is exactly \eqref{trape}. On the other hand, in the banana coordinates, the EOW brane profile becomes very complicated. For $b_t=0$, we obtain
\be 
{\lambda  z r_h=\sqrt{\left(x-b_x \left(t^2+x^2+z^2\right)\right){}^2+t^2} \sinh \left(r_h\left[\pi\pm \tan ^{-1}\left(\frac{t}{x-b_x \left(t^2+x^2+z^2\right)}\right)\right]\right)}, \label{bananaEOW}
\ee 
from which we can not solve $z$ in terms of $(t,x)$ explicitly. However, we have a good evidence that the cancellation  proceeds in the banana case. Notice that even though the banana metric is also very complicated, however its determinant is still given by $\sqrt{g}=1/z^3$. This allows us to perform the $z$-integration first, leading to a surface integral at the AdS boundary. This integral, when combined with the counter-term $I_{ct}$
, yields the desired result for the two-point correlation function. Additionally, there is another surface integral at the EOW branes, which cancels with $I_{\text{EOW}}$.
This cancellation is quite general and can be understood from the fact that adding EOW branes to the system does not alter the ADM mass. Since the EOW branes considered here do not introduce any physical boundaries in the CFT, there are no contributions from Weyl anomalies. The only effect of the EOW brane is to hide the horizon. Therefore, it removes the associated Wald entropy functional \cite{Wald:1993nt} from the on-shell action, which renders the on-shell action to be equal to the energy instead of free energy. In Appendix \ref{appendix:bananaEOW}, we first demonstrate the EOW brane \eqref{bananaEOW} indeed has a profile of a banana and then  explicitly show the cancellation of the surface integrals along the EOW brane by zooming in each section of the banana geometry. 

%%%%%%%%%%%%%%%%%%%%%%%%%%
\section{Spacetime Bananas with Spins}
%%%%%%%%%%%%%%%%%%%%%%%%%%
\label{section:spins}
\renewcommand{\theequation}{4.\arabic{equation}}
\setcounter{equation}{0}
In this section, we consider a rotating BTZ black hole and follow the same strategy to construct a spinning spacetime banana. Our motivation here is to compute the two-point function of huge spinning operators from the information in the dual gravity. In a CFT in $d\geq 3$ Euclidean dimensions, the two-point function of spinning operators takes the general form \cite{Lauria:2018klo}
\be 
G_2(x_1,x_2)\sim \frac{H_{12}^\ell}{|x_{12}|^{2\Delta}}, \label{G2}
\ee 
where $H_{12}^\ell$ is a specific function of $x_1^a$ and $x_2^b$,  which labels the tensor structure of the two operators. For example, the two-point function of two primary vector operators is given by
\be 
\langle v_a(x_1)v_b(x_2)\rangle=c_1\frac{w_{ab}}{|x_{12}|^{2\Delta}},\quad w_{ab}=\delta_{ab}-2\frac{(x_{12})_a(x_{12})_b}{x_{12}^2}.
\ee 
However, as demonstrated in \cite{Janik:2010gc} and revisited in Appendix \ref{appendix:string}, the semi-classical string calculation alone can only reproduce the term \eqref{actionspin} associated with the conformal dimension. To account for the spin structure term $H_{12}^\ell$, \cite{Kazama:2011cp,Kazama:2013qsa} proposed introducing a semiclassical wavefunction for each operator inserted on the string worldsheet, subsequently re-expressing these wavefunctions through appropriate action-angle variables by exploiting the system's integrability. Moreover, by demanding that the wavefunctions transform as primary operators under $SL(2,C)_L \times SL(2,C)_R$ isometry transformations, the corresponding two-point function becomes attainable. This implies that the pure Einstein gravity on-shell action alone is insufficient to generate the spin structure term. Notably, the Einstein gravity on-shell action is manifestly real, while the spin structure term in 2D CFTs is purely imaginary. In subsequent sections, we shall employ the strategy in \cite{Janik:2010gc} to derive the semi-classical gravitational action for computing $\frac{1}{|x_{12}|^{2\Delta}}$ and subsequently proposed a potential approach to recover the spin structure function. 
\\We start from the semi-classical solution of the gravity theory, namely a rotating BTZ black hole solution \cite{Banados:1992gq,Banados:1992wn}
\be\label{eq:sbtz}
ds^2=\frac{(r^2-R_+^2)(r^2+R_-^2)}{r^2}d\t^2+\frac{r^2}{(r^2-R_+^2)(r^2+R_-^2)}dr^2+r^2 \( d\phi-\frac{R_+R_-}{r^2}d\t \)^2,
\ee 
where the outer(inner) horizon radii $R_+(R_-)$ are related to the two periods $(\beta, \theta)$ of the coordinates, 
\bea 
&&(\phi,\t)\sim (\phi+2\pi,\t),\\
&&(\phi,\t)\sim (\phi+\theta,\t+\beta),
\eea 
via
\be 
\beta=\frac{2\pi R_+}{R_+^2+R_-^2},\quad \theta=\frac{2\pi R_-}{R_+^2+R_-^2}.
\ee 
It is straightforward to compute the AdS Einstein-Hilbert action of this solution
\be 
I_{\text{AdS-EH}}=-\frac{c\pi^2}{3} \( \frac{ \beta}{\beta^2+\theta^2} \) =\beta M_{BH}+J_{BH}\theta-S_{BH}=F_{\text{Gibbs}}, \label{actionrbtz}
\ee 
where the mass $M_{BH}$, angular momentum $J_{BH}$ and entropy of the black hole are given by 
\bea 
S_{BH}&=&\frac{c2\pi }{6}R_+=\frac{2\pi^2c}{3} \( \frac{\beta}{\beta^2+\theta^2} \),\\
 M_{BH}&=&\frac{c2\pi }{24\pi}(R_+^2-R_-^2)=\frac{c\pi^2}{3}\frac{(\beta^2-\theta^2)}{(\beta^2+\theta^2)^2},\label{eq:mbh}\\
  J_{BH}&=&\frac{c2\pi}{12\pi}R_-R_+=\frac{2c\pi^2}{3}\frac{\beta\theta}{(\beta^2+\theta^2)^2}.\label{eq:jbh}
\eea 
The problem is the same as the one in the string calculation, namely the on-shell action is equal to the free energy, and hence it is not proportional to the conformal dimension. The proper action is the so-called micro-canonical action. In a gravity theory, the micro-canonical action is obtained by adding a boundary term \cite{Brown:1992bq} which changes the boundary condition from fixed metric components to fixed energy density and momentum density. It should be emphasized that in \cite{Brown:1992bq} the boundary term is added at the spacetime boundary. Here, motivated by the banana proposal, we will add this term at the stretched horizon. Actually, when the momentum density is zero (for a non-rotating black hole), this boundary term is exactly the GHY term. A similar idea has also been explored in \cite{Chua:2023srl} for defining the black hole wave-functions with fixed energy and momentum. To be explicit, the on-shell action we propose is  the modified micro-canonical action: 
\be 
I_{\text{mic}}=\underbrace{-\frac{1}{16\pi G_N}\int d^3x\sqrt{g}(\mathcal{R}+2)}_{\text{AdS Einstein-Hilbert}}\underbrace{-\frac{1}{8\pi G_N}\int_{\partial_\text{AdS}}\sqrt{h}(\mathcal{K}-1)}_{I_{\text{GHY}}(\partial_{\text{AdS}})+I_{ct}}{+I_{\text{bdy}}}(\text{stretched horizon}),\label{eq:action2}
\ee 
where $I_{bdy}$ denotes an additional boundary term localized at the stretched horizon, to be defined in the following section in eq. \eqref{micbdy}.
\subsection{Micro-canonical action}
We will consider a black hole in general $d$ dimensions. To define this boundary term, we need to set up a double foliation of the spacetime \cite{Brown:1992bq}. In our case, we are interested in the stretched horizon, so we first foliate the black hole solution \eqref{eq:sbtz} by surfaces of constant  $r$. Then, on each constant-$r$ surface $\Sigma_r$, we define the induced metric $h_{ij}$, extrinsic curvature $K_{ij}$ and outward-pointing unit normal vector $u^\mu$. We further do an ADM decomposition on $\Sigma_r$:
\be 
h_{ij}dx^i dx^j=N^2d\tau^2+\sigma_{AB}(dy^A+N^A d\tau)(dy^B+N^B d\tau),
\ee 
which indicates the foliation of $\Sigma_r$ by constant-$\tau$ surface $\Gamma_\tau$. Therefore, $\sigma_{AB}$ is the induced metric on $\Gamma_\tau$ and $n_i=N\partial_i \tau$ is the unit normal vector.
Then, we can define the canonical momentum conjugate to $h_{ij}$ on $\Sigma_r$ as follows
\be 
P^{ij}=\frac{\sqrt{h}}{16\pi G_N}(\mathcal{K}^{ij}-h^{ij}\mathcal{K}).
\ee 
Moreover, the associated energy density $q$ and momentum density $j_A$ are given by
\be 
j^i=-\frac{2P^{ij}n_j}{\sqrt{h}}=qn^i+\sigma_A^i j^A,
\ee 
where
\be 
q=\frac{1}{8\pi G_N}\mathcal{K}^{ij}\sigma_{ij},\quad j_A=-\frac{1}{8\pi G_N}\sigma_A^i \mathcal{K}_{ij}n^j,\quad \sigma^i_A=\frac{\partial x^i}{\partial y^A}.
\ee 
The boundary term for changing the canonical ensemble to micro-canonical ensemble is
\be
I_{\text{bdy}} =-\frac{1}{8\pi G_N}\int_{\Sigma_r}d^{d-1}x\sqrt{h}\mathcal{K}+\int_{\Sigma_r}d^{d-1}x\sqrt{\sigma}j_AN^A = I_{\text{GHY}}+I_{J}, \label{micbdy}
\ee
which reduces to the GHY term when $j_A=0$. For our rotating BTZ black hole, we find that the extrinsic curvature is \footnote{Assume that the normal vector points to the center.}
\be 
\mathcal{K}_{rr}=0,\quad \mathcal{K}_{\t\t}=-\sqrt{(r^2+R_-^2)(r^2-R_+^2)}=\mathcal{K}_{\phi\phi},
\ee 
and the normal vector is
\be 
n_i=\delta_{i\t}\frac{\sqrt{(r^2+R_-^2)(r^2-R_+^2)}}{r}.
\ee 
Therefore, the momentum density is as follows
\be
j_\phi=-\frac{1}{8\pi G_N}\delta^i_\phi \mathcal{K}_{ij}n^j=\frac{1}{8\pi G_N} \( \frac{R_-R_+}{r} \).
\ee  
Substituting it into $I_J$ at the outer horizon and using eq. \eqref{eq:jbh} and $N^\phi =-R_-R_+/r^2$, we get
\be 
I_J=\int dt \int d\phi \sqrt{g_{\phi\phi}}j_\phi N^\phi=-\beta 2\pi R_+\frac{1}{8\pi G_N} \( \frac{R_-R_+}{R_+} \) \( \frac{R_-R_+}{R_+^2} \) =-\theta J_{BH},
\ee 
which exactly cancels the second term in eq. \eqref{actionrbtz} as desired. A few technical remarks are in order : First, the ADM decomposition of $\Sigma_r$ is not unique and for different decompositions the value of  $I_J$ is also different. For example, if one chooses a foliation by constant-$\phi$ surfaces, the resulting $I_J$ will be equal to
\be 
-\theta J_{BH} \( \frac{\beta}{\theta} \)^2.
\ee 
Second, it is crucial to do the constant-$\tau$ foliation because the ADM mass and angular momentum of the black hole are defined by the constant-$\tau$ foliation of the AdS boundary. Therefore, only by doing the same ADM decomposition, the charge densities defined on the stretched horizon are compatible with the ones defined on the AdS boundary. Third, in the practice of the computation, sometimes it is more convenient to lift all tensors defined on the surface $\Sigma_r$ to the ones defined in the total spacetime as follows
\be \label{eq:mutensor}
\mathcal{K}_{\mu\nu}=\nabla_{\mu}u_\nu,\quad j^\mu=-\frac{1}{8\pi G_N}\sigma^\mu_A \mathcal{K}_{\mu\nu}n^\nu,\quad \sigma_A^\mu=\frac{\partial x^\mu}{\partial y^A}.
\ee 

\subsection{GtP transformation of rotating BTZ}
\begin{figure}[h]
\begin{center}
  \includegraphics[scale=0.4]{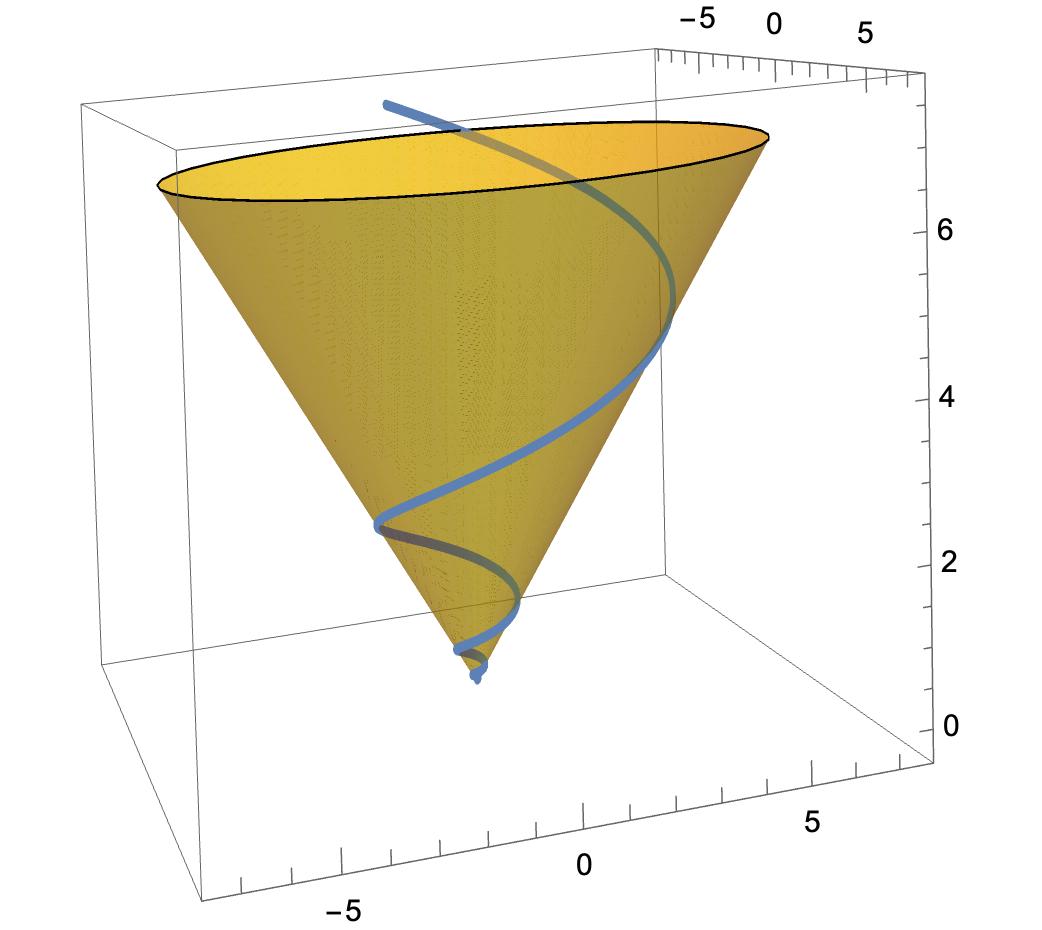}
  \caption{Under the GtP transformations, a spinning circle is mapped to a helix in the cone geometry. Here we set $r_0=1,a=\frac{1}{2}$.}\label{helix}
  \end{center}
\end{figure}

Applying the GtP map \eqref{eq:goP} to the metric \eqref{eq:sbtz}, we get the spinning cone metric
\be 
ds^2= g_{RR} dR^2 + g_{ZZ} dZ^2 + g_{\Theta \Theta} d \Theta^2 + 2 g_{RZ} dR dZ + 2 g_{R \Theta} dR d\Theta +2 g_{Z \Theta} dZ d\Theta,
\ee 
where
\bea
g_{RR} &=& \frac{R^2}{(R^2 + R_{-}^2 Z^2) (R^2 - R_{+}^2 Z^2)}+ \frac{R^2 (R^2 + (R_-^2 - R_+^2) Z^2)}{Z^2 (R^2 + Z^2)^2},
\cr && \cr
g_{Z Z} &=& \frac{R^2}{(R^2 + Z^2)^2} + \frac{(R_-^2 - R_+^2) Z^2}{(R^2 + Z^2)^2} + \frac{R^4}{Z^2 (R^2 + R_-^2 Z^2) (R^2 - R_+^2 Z^2)},
\cr && \cr 
g_{\Theta \Theta} &=& \frac{R^2}{Z^2},
\cr && \cr
g_{R Z} &=& \frac{R^2 (R_-^2 - R_+^2) Z}{(R^2 + Z^2)^2} + \frac{R^3}{Z} \left( \frac{1}{(R^2 + Z^2)^2} + \frac{1}{(R^2 + R_-^2 Z^2) (R_+^2 Z^2 - R^2)}\right),
\cr && \cr 
g_{R \Theta} &=& - \frac{R_- R_+ R}{(R^2 + Z^2)},
\cr && \cr
g_{Z \Theta} &=& - \frac{R_- R_+ Z}{(R^2 + Z^2)}.
\eea 
To visualize the spinning effect, we can look at a rotating circle  described by $r=r_0,\phi=a\tau$. It is straightforward to verify that its path will be mapped to a helix in the cone geometry as shown in Figure \ref{helix}.
The calculation of the AdS Einstein-Hilbert on-shell action is very similar to the non-spinning case. We obtain
\bea  
&&I_{\text{AdS}}=\underbrace{\frac{1}{4G_N}\int_{\epsilon_r}^{\Lambda_r}   dR \, \(\frac{1}{\epsilon^2}-\frac{R_+^2}{R^2}\)}_{\text{AdS Einstein-Hilbert}}\underbrace{-\frac{1}{4G_N}\int_{\epsilon_r}^{\Lambda_r}  RdR\(\frac{1}{\epsilon^2}+\frac{  \( -1+R_-^2-R_+^2 \) }{2R^2}\)}_{I_{GHY}(\partial_{\text{AdS}})+I_{ct}}\nn
&&\qquad\,=-\frac{\( R_-^2+R_+^2-1 \)}{8G_N}\log \( \frac{\Lambda_r}{\epsilon_r} \)
\nn
&&\qquad\,
=\(M_{BH}+\frac{c}{12}+ \frac{\theta J_{BH}}{\beta}-\frac{S_{BH}}{\beta}\)\log \( \frac{\Lambda_r}{\epsilon_r} \),\label{spinAdS}
\eea  
where we have rewritten the result in terms of the charges of the black hole. 
\subsection{The stretched horizon}
In the cone coordinates, the stretched horizon is defined as follows
\be \label{spinstretch}
R=ZR_+(1+\epsilon_s),
\ee 
and the (unnormalized) normal vector can be computed in a similar way as \eqref{eq:normalstretched}, yielding
\be 
V_{R}=-\frac{1}{(1+\epsilon_s)R_+},\quad V_{\Theta}=0,\quad V_{Z}=1.
\ee  
Therefore, according to the definitions of the extrinsic curvature \eqref{eq:mutensor}, we find 
\bea 
&&\mathcal{K}_{RR}=-\frac{\sqrt{2\epsilon_s}R_+^5\sqrt{R_-^2+R_+^2}}{R^2(1+R_+^2)^2},\quad \mathcal{K}_{RZ}=-\frac{\sqrt{2\epsilon_s}R_+^4\sqrt{R_-^2+R_+^2}}{R^2(1+R_+^2)^2},\\
&&\mathcal{K}_{\Theta\Theta}=-\sqrt{2\epsilon_s}R_+\sqrt{R_-^2+R_+^2},\quad \mathcal{K}_{ZZ}=-\frac{\sqrt{2\epsilon_s}R_+^3\sqrt{R_-^2+R_+^2}}{R^2(1+R_+^2)^2},
\eea 
in the leading order of $\epsilon_s$. Setting $Z=R/(R_+(1+\epsilon_s))$, we find the induced metric on the stretched horizon is given by
\bea 
h_{\mu\nu}dx^\mu dx^\nu= \frac{\left( R_-^2+\epsilon_s(2+\epsilon_s)R_+^2 \right)}{R^2}dR^2- \( \frac{ 2 R_-R_+}{R} \) dRd\Theta+ R_+^2 (1+\epsilon_s)^2 d\Theta^2.
\eea  
As we discussed in last section, different ADM decompositions of the stretched horizon will give different results for the boundary action. Therefore, to be compatible with the black hole charges, we should not change the ADM decomposition, namely the constant-$\tau$ decomposition. Then, in the cone coordinates, the GtP map \eqref{eq:goP} implies we should consider a surface with fixed value of $R^2+Z^2$. Considering that on the stretched horizon $Z$ is related to $R$ via \eqref{spinstretch}, the proper ADM decomposition is 
\be 
h_{\mu\nu}dx^\mu dx^\nu=\frac{\epsilon_s(2+\epsilon_s) \( R_-^2+(1+\epsilon_s)^2R_+^2 \)}{(1+\epsilon_s)^2R^2}dR^2+ R_+^2 (1+\epsilon_s)^2 \( d\Theta-\frac{R_-}{(1+\epsilon_s)R}dR \)^2.
\ee 
Having fixed the ADM decomposition, we find
\bea  
&&\sqrt{h}=\frac{R_+ \sqrt{2\epsilon_s}\sqrt{R_-^2+R_+^2}}{R},\quad \mathcal{K}=-\frac{\sqrt{R_-^2+R_+^2}}{\sqrt{2\epsilon_s}R_+},\\
 &&j_\Theta=\frac{R_-}{8\pi G_N},\quad N^\Theta=-\frac{R_-}{(1+\epsilon_s)R}.
\eea  
Substituting \eqref{sGHY} and \eqref{sNJ} into \eqref{micbdy} gives 
\bea  
I_{\text{GHY}}(\text{stretch})&=&\frac{\left( R_-^2+R_+^2 \right)}{4 G_N}\int_{\epsilon_r}^{\Lambda_r} \frac{dR}{R}= \( \frac{S_{BH}}{\beta} \) \log \( \frac{\Lambda_r}{\epsilon_r} \),\label{sGHY}\\
I_J&=&-\frac{R_-^2}{4G_N}\int_{\epsilon_r}^{\Lambda_r} \frac{dR}{R}=- \( \frac{\theta J_{BH}}{\beta} \) \log \( \frac{\Lambda_r}{\epsilon_r} \)\label{sNJ}.
\eea  
Therefore, after combing the Einstein-Hilbert action \eqref{spinAdS} with \eqref{sGHY} and \eqref{sNJ}, we end up with the desired micro-canonical action
\be 
I_{\text{mic}}=I_{\text{AdS}}+I_{\text{GHY}}(\text{stretch})+I_J=\(M_{BH}+\frac{c}{12}\)\log \( \frac{\Lambda_r}{\epsilon_r} \) =(h+\bar{h})\log \( \frac{\Lambda_r}{\epsilon_r} \),
\ee 
where $h$ and $\bar{h}$ and chiral and anti-chiral conformal dimensions  defined by
\be 
M_{BH}+\frac{c}{12}=h+\bar{h},\quad J_{BH}=h-\bar{h}. \label{crbh}
\ee 
After this long calculation of the boundary action \eqref{sGHY} and \eqref{sNJ} on the stretched horizon, now it is clear that the whole point of adding this term is to cancel the surface integral on the horizon coming from the Einstein-Hilbert action when the $z$ variable is integrated. As we proposed in section \ref{section:banana}, we can directly compute the Einstein-Hilbert action of the extrapolated cone geometry to get the following result
\bea  
I_{\text{ex-cone}}&=&\underbrace{\frac{1}{4G_N}\int_{\epsilon_r}^{\Lambda_r}  \,  dR \(\frac{1}{\epsilon^2}\)}_{\text{AdS Einstein-Hilbert}}\underbrace{-\frac{1}{4G_N}\int_{\epsilon_r}^{\Lambda_r}  RdR\(\frac{1}{\epsilon^2}+\frac{\( -1+R_-^2-R_+^2 \)}{2R^2}\)}_{I_{GHY}(\partial_{\text{AdS}})+I_{ct}},\nn
&=&-\frac{1}{4 G_N}\int_{\epsilon_r}^{\Lambda_r} RdR  \frac{\( R_-^2-R_+^2-1 \)}{2 R^2}= \( M_{BH}+\frac{c}{12} \) \log \( \frac{\Lambda_r}{\epsilon_r} \).
\eea

\subsection{EOW brane in spinning cone}
\begin{figure}[h]
\begin{center}
  \includegraphics[scale=0.3]{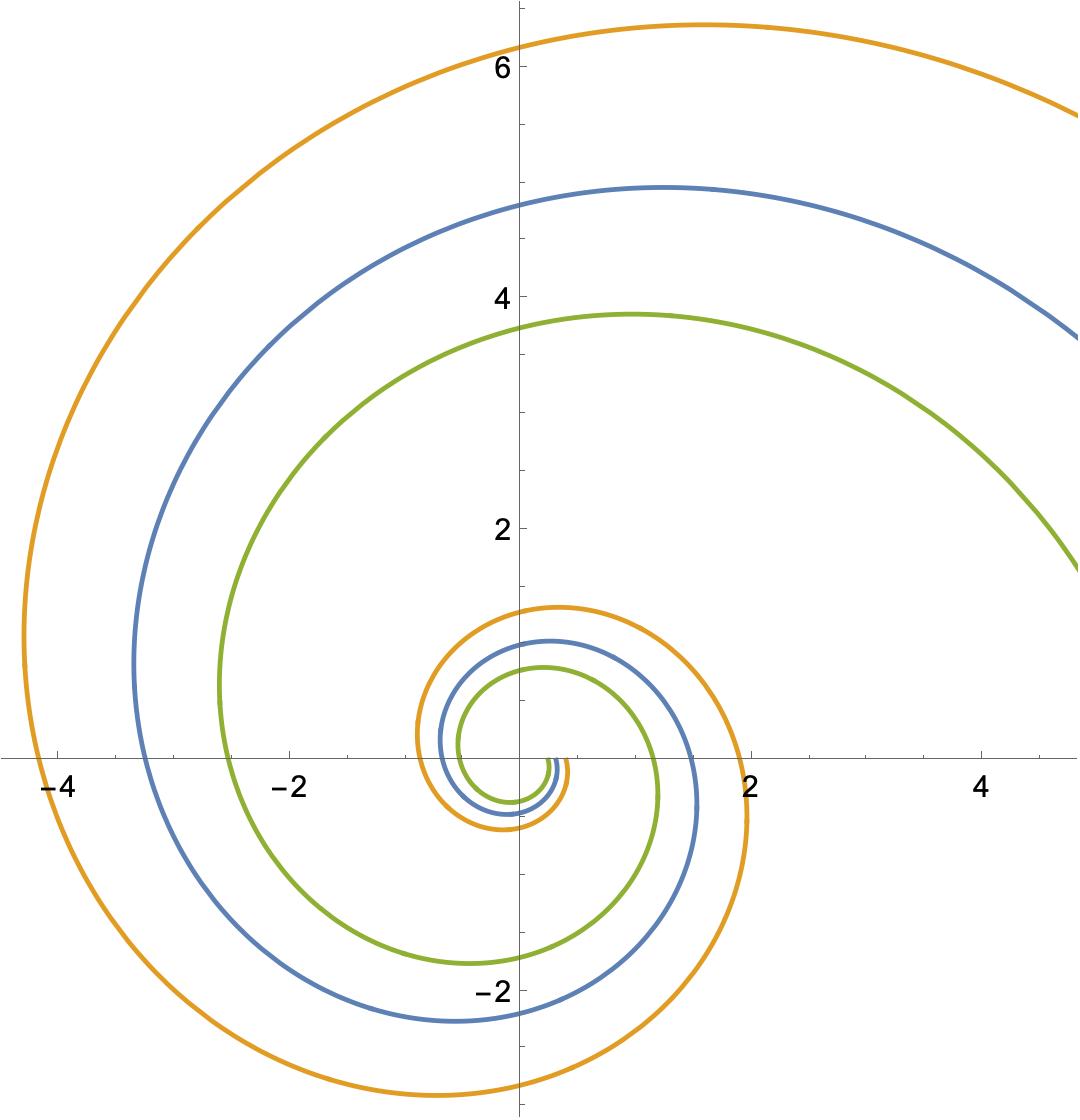}
  \caption{Projecting the EOW brane in the spinning cone coordinates to the AdS boundary gives a bundle of helices described by eq. \eqref{helices}.}\label{helixEOW}
  \end{center}
\end{figure}

As we proposed in section \ref{subsection:onepoint}, the cone geometry can be used to compute the one-point function on a disk by incorporating the EOW branes in the framework of AdS/BCFT. Here we explore this possibility. The time-like EOW brane in the rotating BTZ geometry is described by the profile equation
\be 
\sqrt{\frac{r^2-R_+^2}{R_+^2+R_-^2}}\cos(R_+\tau+R_-\phi)=-\lambda.
\ee 
Moreover, its projection on the AdS boundary is a bundle of lines
\be 
R_+\tau+R_-\phi=\frac{\pi}{2}+k\pi,\quad k\in \mathbb{Z}.
\ee 
Then one can easily verify that the GtP transformation maps them to the a bundle of helices
\be 
R=\exp\(\frac{\frac{\pi(k+1)}{2}-R_-\Theta}{R_+}\),
\label{helices}
\ee 
which we plot in Figure \ref{helixEOW}. 
Therefore, we cannot use the spinning cone geometry to compute the spinning one-point function on a disk. This is not a failure because from a bootstrap argument \cite{Kusuki:2022ozk}, the one-point function of a non-scalar primary operator actually vanishes. Our conclusion is also reminiscent to the holographic argument \cite{Kusuki:2022ozk} of vanishing of spinning one-point function. In \cite{Kusuki:2022ozk}, the authors proposed the construction of a geometry with spinning defect by a twisted identification of wedges in global AdS. Moreover, they argued this twisted identification will cause a singular region around the intersection between a brane and the spinning defect, thus this configuration is forbidden. Our construction does not allow a configuration for computing disk one-point function, either. However, it has a non-trivial prediction that the spinning one-point function does \textbf{not} vanish on a disk with a helical boundary. It will be very interesting to verify this from a CFT calculation.
\subsection{Spinning bananas}
Having figured out the suitable boundary term to add at the stretched horizon,  we can now perform the SCT transformation in \eqref{stp}
%%%%%%%%%%%%%%%%%%%%%%%%%%%%%%%%%%%%%%
\footnote{Notice that for simplicity we set $b_t=0$.}
%%%%%%%%%%%%%%%%%%%%%%%%%%%%%%%%%%%%%%
to obtain the spinning banana spacetime. As we proposed and showed in the example of the spinning cone geometry, we can directly compute the Einstein-Hilbert action of the extrapolated banana geometry, which is also equivalent to computing the micro-canonical action. Similar to \eqref{excone}, we only need to know the determinant of the induced metric at the AdS boundary which is as follows
\be 
\sqrt{h}=\frac{1}{\epsilon^2}-\frac{1+R_+^2-R_-^2}{2(t^2+x^2)(b_x^2t^2+(b_xx-1)^2)}.
\ee 
Then the micro-canonical on-shell action is given by
\bea  
I_{\text{mic}}&=&\frac{\( 1+R_+^2-R_-^2 \)}{16\pi G_N}\int dtdx\frac{1}{(t^2+x^2)(b_x^2t^2+(b_xx-1)^2)}\\
&=&\frac{\( 1+R_+^2-R_-^2 \)}{8 G_N}\log\left|\frac{1/b_x}{\epsilon}\right|^2=\Delta\log\left|\frac{1/b_x}{\epsilon}\right|^2,
\eea  
where we have used \eqref{eq:mbh},\eqref{crbh} and the result \eqref{integral}. As promised, the partition function is equal to the two-point function
\be 
e^{-I_{\text{mic}}}=\langle\mathcal{O}(0)\mathcal{O}(b_x^{-1})\rangle=\(\frac{\epsilon}{1/b_x}\)^{2\Delta}.
\ee 
\subsection{Towards the spin structure function}
As mentioned at the beginning of this section, to reproduce the spin structure function, we have to add a new purely imaginary term to the gravity action. A proposal from a string theory calculation suggests considering incorporating terms corresponding to the operators themselves. Our proposal is motivated by the following argument. When inserting an operator with moderate conformal dimension at the global AdS$_3$ center, its backreaction generates the conical AdS$_3$ geometry. The resulting conical singularity directly stems from the operator's presence. As the conformal dimension increases, the backreaction ultimately induces black hole formation that encapsulates the operator, leaving a (conical) singularity. This motivates us to mimic the spinning effect of the inserted operator by the twist of the geodesic connecting them. For a spinning particle (or anyon), the action associated with the twist is given by \cite{Castro:2014tta}
\be 
S_{\text{twist}}=\im J \eta,
\ee 
where $J$ is the angular momentum and $\eta $ is the twist along the geodesic. To define the twist, we introduce parallel transported frames $v^\alpha,n_1^\alpha,n_2^\alpha$ along the geodesic, satisfying 
\be 
v^\alpha=\frac{dX^\alpha}{d\tau},\quad \nabla n_1=\nabla n_2=0,\quad n_1\cdot n_2=n_i\cdot v=0,\quad n_i\cdot n_i=1,
\ee 
where we choose the affine parameter $\tau$ such that the tangent vector is normalized and the differential operator is defined as
\be 
\nabla V^\mu=\frac{dV^\mu}{d\tau}+\Gamma_{\lambda\rho}^\mu \frac{dX^\rho(\tau)}{d\tau}V^\lambda.
\ee 
Note that the vectors $n_i$ are defined up to a rotation
\be 
n_1\to \cos(\alpha)n_1+\sin(\alpha)n_2,\quad n_2\to -\sin(\alpha)n_1+\cos(\alpha)n_2,
\ee 
where $\alpha$ is some constant. Then we can define the twist as $\eta=\arccos \left( n_1(\tau_i)\cdot n_1(\tau_f) \right)$. To remove the ambiguity due to the choice of $\alpha$, we adopt a strategy that is similar to the one used in the string worldsheet calculation, namely starting from some reference then applying the isometry transformation. In the \Poincare coordinates where
\be 
ds^2=\frac{dz^2+dt^2+dx^2}{z^2},
\ee 
the reference geodesic is given by
\be 
z=e^{\tau},\quad x=t=0.
\ee 
Moreover, the tangent vector is 
\be 
v^z= e^{\tau},\quad v^t=v^x=0.
\ee 
Due to the rotational symmetry, the other two frame vectors are
\be 
n_1=\partial_t,\quad n_2=\partial_x.
\ee 
The twist vanishes since $n_1$ is a constant vector. Next, we apply the SCT transformation
\bea 
&&\tilde{t}=\frac{t+b_t(t^2+x^2+z^2)}{1+2(b_x x+b_t t)+(b_x^2+b_t^2)(t^2+x^2+z^2)},\\
&&\tilde{x}=\frac{x+b_x(t^2+x^2+z^2)}{1+2(b_x x+b_t t)+(b_x^2+b_t^2)(t^2+x^2+z^2)},\\
&&\tilde{z}=\frac{z}{1+2(b_x x+b_t t)+(b_x^2+b_t^2)(t^2+x^2+z^2)},
\eea 
which maps $(z(\tau_i),t(\tau_i),x(\tau_i))=(0,0,0)$ and $(z(\tau_f),t(\tau_f),x(\tau_f))=(\frac{1}{\epsilon},0,0)$ to 
\bea 
(\tilde{z}(\tau_i),\tilde{t}(\tau_i),\tilde{x}(\tau_i)) &=& (0,0,0), %\quad
\cr && \cr
(\tilde{z}(\tau_f),\tilde{t}(\tau_f),\tilde{x}(\tau_f)) &=& \left( \frac{\epsilon}{b_x^2+b_t^2+\epsilon^2},\frac{b_t}{b_x^2+b_t^2+\epsilon^2},\frac{b_x}{b_x^2+b_t^2+\epsilon^2} \right).
\eea 
Correspondingly, the vector $n_1$ are mapped to
\be 
n_1(\tau_i)=\partial_t,\quad n_1(\tau_f)= \left( \frac{b_x^2-b_t^2}{b_t^2+b_x^2} \right) \partial_t- \left( \frac{2b_tb_x}{b_t^2+b_x^2} \right) \partial_x,
\ee 
which leads to
\be 
n_1(\tau_i)\cdot n_1(\tau_f)=\cos(2\theta),\quad b_x\equiv R\cos\theta,\quad b_t \equiv R\sin\theta,\to \eta=2\theta =-\im \log \left( \frac{b_x+\im b_t}{b_x-\im b_t} \right),
\ee 
and
\be 
e^{-\im J \eta}=\(\frac{b_x-\im b_t}{b_x+\im b_t}\)^J,
\ee 
as expected. A few comments are in order. In \cite{Castro:2014tta}, a similar action
\be 
S_{EE}=\int_C ds\(m \sqrt{g_{\mu\nu}\dot{X}^\mu \dot{X}^\nu}+\im J\tilde{n}\cdot \nabla n\),
\ee 
is proposed to describe a spinning particle, where ${n}$ and ${\tilde{n}}$ are two unit vectors perpendicular to the trajectory (this is ensured by the constraint term whose explicit expression is not relevant for us.) In particular, the variation of this action leads to the Mathisson-Papapetrou-Dixon equations. Furthermore, it is demonstrated that this on-shell action correctly reproduces the two-point function of two spinning operators. However, to reproduce this desired result, a special gauge has to be imposed. To show this subtlety,  following \cite{Castro:2014tta} let us consider the particular geodesic 
\bea  
z=\frac{1}{2} \text{sech}(k \tau ),\quad t=\frac{1}{2} \sin (\theta ) (\tanh (k \tau )+1),\quad x=\frac{1}{2} \cos (\theta ) (\tanh (k \tau )+1),
\eea  
in the \Poincare EAdS metric
\be 
ds^2=\frac{dz^2+dt^2+dx^2}{z^2}.
\ee 
Note that the profile is a half-big-circle
\be 
z^2+ \left( x-\frac{\cos\theta}{2}\right)^2+ \left( t-\frac{\sin\theta}{2} \right)^2=\frac{1}{4}.
\ee 
The geodesic touches the AdS boundary at the  two insertion points:
\be 
w_1=(x_1,t_1)=(0,0),\quad w_2=(x_2,t_2)=(\cos\theta,\sin\theta).
\ee 
The (unnormalized) tangential vector  of the geodesic, defined as $v^\mu=\frac{dX^\mu}{d\tau}$ is
\bea 
v^z=-\frac{1}{2} k \tanh (k \tau ) \text{sech}(k \tau ),\quad v^t=\frac{1}{2} k \sin (\theta ) \text{sech}^2(k \tau ),\quad v^x=\frac{1}{2} k \cos (\theta ) \text{sech}^2(k \tau ). \;\;\;
\eea 
One can easily check that
\be 
\nabla v=0.
\ee 
Next let us solve the parallel transported vectors which are orthogonal to $v$, defined by
\be 
\nabla q_1=\nabla q_2=0,\quad q_1\cdot q_2=q_i\cdot v=0,\quad q_i\cdot q_i=1.
\ee 
The solution is 
\bea  
&&q_1^z=0,\quad q_1^t=\frac{1}{2}\cos\theta \,\text{sech}(k \tau ),\quad q_1^x=-\frac{1}{2}\sin\theta\, \text{sech}(k \tau ),\\
&&q_2^z=\frac{1}{2} \text{sech}^2(k \tau ),\quad q_2^t=\frac{1}{2} \sin (\theta ) \tanh (k \tau ) \text{sech}(k \tau ),\quad q_2^x=\frac{1}{2} \cos (\theta ) \tanh (k \tau ) \text{sech}(k \tau ). \nn
\eea  
Note that the choice of $q_1,q_2$ are not unique but can be fixed up to a rotation
\be 
q_1\to \cos(\alpha)q_1+\sin(\alpha)q_2,\quad q_2\to -\sin(\alpha)q_1+\cos(\alpha)q_2.
\ee 
Expanding $n$ and $\tilde{n}$ as
\bea 
&&n=\cos(\eta(\tau))q_1+\sin(\eta(\tau))q_2,\\
&&\tilde{n}=-\sin(\eta(\tau))q_1+\cos(\eta(\tau))q_2,
\eea 
then we exactly find that the following expression
\be 
\int \tilde{n}\cdot \nabla n=\int\dot{\eta}=\eta(x_f)-\eta(x_i).
\ee 
To solve the twist $\eta$, one should fix $\alpha$ and then solve the equation of motion of $n$. However, \cite{Castro:2014tta} suggests an ad hoc method to find $\eta$ directly. The proposal of \cite{Castro:2014tta} is that if we fix $\alpha=0$ then choose $n_1(\text{initial})=n_1(\text{final})=\partial_t$, then we can end up with
\be 
\eta
=-\im \log \left( \frac{n_f\cdot q_1+\im n_f \cdot q_2}{n_i\cdot q_1+\im n_i \cdot q_2} \right) =-\im \log \left( \frac{e^{\im \theta}}{e^{-\im \theta}} \right) =-\im \log \left( \frac{w_1-w_2}{\bar{w}_1-\bar{w}_2} \right),
\ee 
such that
\be 
e^{-\im J \eta }=\(\frac{\bar{w}_1-\bar{w}_2}{w_1-w_2}\)^J.
\ee 
The choice $n_1(\text{initial})=n_1(\text{final})=\partial_t$ is argued by comparing with the CFT computation, but the reason for choosing the choice $\alpha=0$ is not clear except for rendering the desired results.

%%%%%%%%%%%%%%%%%%%%%%%%%%
\section{Discussion}
\label{section:discussion}
%\label{section:spins}
\renewcommand{\theequation}{5.\arabic{equation}}
\setcounter{equation}{0}
%%%%%%%%%%%%%%%%%%%%%%%%%%
In this final section, we first summarize the results of this paper and then discuss some observations and remaining issues for future investigations.
\subsection{Summary}
In this work, we generalized the spacetime banana proposal from various aspects. First, we introduced the EOW branes into this framework, presenting two applications. In the first scenario, the EOW branes intersect with the AdS boundary, resulting in a standard AdS/BCFT setup. Here, we successfully constructed a holographic dual of the one-point function in a BCFT. However, we argued that the current method is inadequate for constructing the holographic dual of the two-point functions. This limitation may reflect the fact that the two-point functions in a BCFT are not universal and cannot be fully determined by conformal symmetry. In the second scenario, we considered a floating EOW brane in the bulk. This EOW brane does not touch the AdS boundary but wraps around the black hole horizon. We found that this EOW brane can play the role of a stretched horizon, as introduced in the original proposal, to project the canonical ensemble onto a specific state. Therefore, our EOW proposal also provides a concrete realization of the idea that EOW branes can be used to describe the micro-states of black holes.
\\Furthermore, we studied the holographic one- and two-point functions of large primary operators with non-zero spins. The dual geometries for these calculations are constructed by starting from a rotating BTZ black hole and applying the GtP and SCT transformations, respectively. We observed that to accurately reproduce the holographic correlators, the addition of the GHY term on the stretched horizon is insufficient. In this context, we proposed a refined version of the micro-canonical action introduced in \cite{Brown:1992bq}. Notably, our proposed action reduces to the one used in the original spacetime banana proposal. Given that the micro-canonical action is quite general and can be defined for any black hole, we anticipate that our proposal can also be extended to the spacetime banana proposal with other charges or matter fields.
\\Finally, to compare the (generalized) spacetime banana proposal, we revisited other existing approaches for computing correlation functions using the on-shell action of gravity. Surprisingly, we found that computing the on-shell action of the gravity solution in the FG gauge yields an incorrect result. We argued that this discrepancy arises from the fact that the FG gauge does not cover the entirety of spacetime. In an upcoming paper \cite{wedge}, we will present additional examples.
%%%%%%%%%%%%%%%%%%%%%%%%%%
\subsection{Twist operator vs. massive particle}
In Appendix \ref{appendix:particle}, we have shown that the on-shell action of the conical geometry in the cone or banana coordinates is different from the one in its corresponding FG gauge by a minus sign and this discrepancy can be resolved by including the on-shell partition function of a missing region in the FG gauge. In Appendix \ref{appendix:twist}, we also show that the on-shell partition function of the \Banados solution, which is supposed to be equal to the two-point function of the twist operators, is actually equal to the inverse of that. Combining these two facts indicates that to produce the two-point function of twist operators, we should complement the \Banados solution \eqref{twistmetric} with other missing part of the spacetime. 
\\First, we want to point out that there is an often overlooked difference between the \Banados solution \eqref{twistmetric} with \eqref{onetwist} corresponding to a twist operator insertion and the \Banados solution \eqref{one-point-LC-FG} with \eqref{oneconical} corresponding to a heavy particle operator insertion. They seem to be equal to each other, if we identify $a^2$ with $1/q^2$. However, they are different entities. To distinguish them, we rename the $u$ coordinate in \eqref{twistmetric} as $u_t$ and the one in \eqref{one-point-LC-FG} as $u_m$. By definition, the metric $du_m d\bar{u}_m$ is smooth while the metric $du_t d\bar{u}_t$ has a conical singularity at the center (going around the center takes an angle of $2\pi q$). Interestingly, we can use \eqref{FG coordinate-transformation} and the GtP map to transform \eqref{twistmetric}
back to the global coordinates. Of course, we will get
\be\label{nonstandard}
ds^2= \( r^2+\frac{1}{q^2} \)d\tau^2+\frac{dr^2}{\( r^2+\frac{1}{q^2} \)}+r^2d\theta^2,\quad \theta\sim 2\pi q+\theta,
\ee 
which turns out to be the smooth global AdS$_3$ in a little unfamiliar coordinates! To summarize what we have just found, there exists a new interpretation of the \Banados solution \eqref{twistmetric}. It can be constructed in the following way: we start from the global AdS$_3$ spacetime in a non-standard coordinates \eqref{nonstandard} followed by the GtP transformation \eqref{gtp2} and in the end rewrite the result in the FG gauge. Now we understand that the \Banados solution \eqref{twistmetric} misses the region inside a cone defined in the cone coordinates. 
However, the interpretation may look a little weird at first glance. How does the smooth global AdS$_3$ geometry end up with a non-trivial \Banados solution? The secret is that when we transform to the FG gauge, we also change the AdS cut-off surface. In the FG gauge, if we choose the conventional cut-off $z=\epsilon$, the boundary metric becomes singular which is the sign of operator insertions. As emphasized in \cite{Abajian:2023jye}, the whole point of the GtP transformation is to introduce a convenient but non-conventional cut-off surface $Z=\epsilon$ for us to compute the CFT correlation functions.
In the second spacetime banana paper \cite{Abajian:2023bqv}, the authors reformulated the computation in a Liouville theory such that the non-triviality of the cut-off can be uniformly encoded in a Liouville field.
\subsection{A spacetime banana in dS}
Is there a specific physical meaning related to the region inside the spacetime cone or banana? Looking back to the Euclidean BTZ geometry, its interior is naturally described by a metric
\be 
ds^2=\( r_h^2-r^2 \)d\tau^2+\frac{dr^2}{\( r_h^2-r^2 \)}+r^2d\phi^2,
\ee  
which describes the geometry of the Euclidean de Sitter black hole. In three dimensions, there is only one horizon at $r=r_h\equiv\sqrt{1-8G_NE}$ and the smoothness condition of the horizon imposes the periodic condition
\be 
\tau\sim \tau+\frac{2\pi}{r_h}.
\ee  
Similar to the BTZ black hole, there is a conical singularity at $r=0$ with the deficit angle $\delta=2\pi(1-r_h)$. The GtP transformation will map it to a de Sitter cone. The Einstein-Hilbert action of this cone is given by 
\be 
I_{\text{dS-Cone}}=-\frac{1}{16\pi G_N}\int  \sqrt{g}(\mathcal{R}-2)=-\frac{r_h^2}{4G_N}\log \( \frac{\Lambda_r}{\epsilon_r} \) =2 \( E-\frac{c}{12} \) \log \( \frac{\Lambda_r}{\epsilon_r} \).
\ee 
Furthermore, by applying the SCT map we can obtain a de Sitter banana. The on-shell action can be similarly computed, the result is as follows
%%%%%%%%%%%%%%%%%%%%%%%%%%%%%%
\footnote{Here for simplicity we set $b_t=0$.}
%%%%%%%%%%%%%%%%%%%%%%%%%%%%%%
\be 
I_{\text{dS-Banana}}=2 \( E-\frac{c}{12} \) \log\left|\frac{1/b_x}{\epsilon}\right|^2.
\ee 
Therefore, we find that the partition function of the de Sitter spacetime banana is equal to the modular square of the two-point function
\be 
\frac{e^{-I_{\text{ds-Banana}}(E)}}{e^{-I_{\text{ds-Banana}}(E=0)}}=\(\frac{\epsilon}{b_x^{-1}}\)^{4\Delta}=|\langle \mathcal{O}(0)\mathcal{O}(b_x^{-1})\rangle|^2,\quad E\equiv \Delta.
\ee 
This is reminiscent of the result in the wormhole/ensemble CFT duality \cite{Chandra:2022bqq} which proposes that the wormhole partition function is equal to modular square of the correlation functions in Liouville CFT
\be 
e^{-I_{\text{wormhole}}}\approx |G_L(z_i,\bar{z}_i)|^2. \label{wormhole}
\ee 
In fact, our de Sitter banana looks very similar to the generalized Maldacena-Maoz wormhole geometry \cite{Chandra:2022bqq}. We conjecture that this is true in general: the interior of an $n$-point geometry, which is a generalization of a spacetime banana, can be described by a de Sitter $n$-point geometry; The on-shell action of the de Sitter $n$-point geometry is equal to the modular square of the $n$-point function up to some normalization. One possible explanation of the modular square is that the physical operator in the de Sitter space is a product of two operators in the AdS space \footnote{We thank Cheng Peng for this suggestion.}. For example, in the de Sitter holography proposed in \cite{Narovlansky:2023lfz}, the 3D de Sitter spacetime is dual to a doubled SYK model and a physical operator is given by the integral of the product of two operators from the copies of the SYK model.
\subsection{The spacetime banana in JT gravity}
In this work, as well as in the original studies \cite{Abajian:2023bqv,Abajian:2023jye}, the discussion focused on the spacetime banana in AdS$_d,\, \text{where } d\geq 3$. We may also consider a similar solution in an AdS$_2$ gravity theory, for example, the JT gravity\cite{Teitelboim:1983ux,Jackiw:1984je}. The Euclidean JT gravity on a manifold $\mathcal{M}$ with the boundary $\partial M$ has the following action
\be 
S=S_{\text{top}}+S_{bulk}+S_{\partial},
\ee 
where the first of these three terms is $-\phi_0 \chi(\mathcal{M})$ with $\phi_0$ a constant and $\chi(\mathcal{M})$ the Euler character of $\mathcal{M}$. The other two pieces are given by
\be 
S_{\text{bulk}}=-\frac{1}{16\pi G_N}\int_{\mathcal{M}}d^2x\sqrt{g}\phi(\mathcal{R}+2),\quad S_{\partial}=-\frac{1}{8\pi G_N}\int_{\partial \mathcal{M}} du \sqrt{h}\phi(\mathcal{K}-1).
\ee 
The equations of motion for the metric and dilaton are given by
\be 
\mathcal{R}+2=0,\quad \(\nabla_\mu\nabla_\nu-g_{\mu\nu}\nabla^2+g_{\mu\nu}\)\phi=0.
\ee 
In the \Poincare coordinates, the solutions are as follows
\be 
ds^2=\frac{d\tau^2+dz^2}{z^2},\quad \phi=\frac{a_1+a_2\tau+a_3(\tau^2+z^2)}{z}.
\ee 
The on-shell action is nearly trivial; however, the non-trivial aspects arise from the non-trivial cut-off, represented by a curve $(\tau(u),z(u))$, where u denotes the intrinsic boundary coordinate. The boundary conditions imposed on this curve are as follows
\be 
ds^2 |_{\partial \mathcal{M}}=g_{uu}du^2=\frac{du^2}{\epsilon^2},\quad \phi|_{\partial \mathcal{M}}=\frac{\phi_r(u)}{\epsilon}.
\ee 
Note that the induced metric on the curve is 
\be 
\frac{\left( z'^2+\tau'^2 \right)}{z^2}du^2
\ee 
and the extrinsic curvature is given by
\be 
\mathcal{K}=1+\epsilon^2\text{Sch}(\tau,u).
\ee 
To construct the 2D spacetime banana let us start with the black hole metric
\be 
ds^2= \( r^2-r_h^2 \)d\tau^2+\frac{dr^2}{\( r^2-r_h^2 \)},\quad \phi=\phi_br,\quad \quad \tau\sim\tau+\beta,\quad \beta\equiv\frac{2\pi}{ r_h}.
\ee 
In the black hole metric, if we choose the cut-off surface at $r=\Lambda$, the corresponding extrinsic curvature is given by
\be 
\mathcal{K}=1+\frac{r_h^2}{2\Lambda^2},
\ee 
thus the on-shell action is 
\be 
I_{\text{JT black hole}}=-\frac{1}{8\pi G_N} \( \frac{\beta \phi_b r_h^2}{2} \) =-\beta H,
\ee 
where $H\equiv \frac{\phi_b r_h^2}{16\pi G_N}$ is the Hamiltonian of the JT gravity \cite{Harlow:2018tqv}. Then, by applying similar GtP transformations:
\be 
r=\frac{|T|}{Z},\quad \tau=\frac{1}{2}\log(Z^2+T^2),
\ee 
we get the JT ``cone" solution
\be 
ds^2=\frac{(T^2+Z^2)^2(ZdT-TdZ)^2+(T^2-r_h^2 Z^2)^2(TdT+ZdZ)^2}{Z^2(T^2+Z^2)^2(T^2-r_h^2 Z^2)},\quad \phi=\frac{\phi_b T}{Z},
\ee 
which only covers the region outside the wedge defined by $|T|=r_h Z$.
The corresponding on-shell action is 
\be 
I_{\text{JT cone}}=-\frac{(1+r_h^2)\phi_b}{16\pi G_N}\int_{\epsilon_t}^{\Lambda_t}\frac{dT}{T}=-\Delta \log \( \frac{\Lambda_t}{\epsilon_t} \),\quad \Delta\equiv \frac{(1+r_h^2)\phi_b}{16\pi G_N},
\ee 
and adding a GHY term at the stretched horizon yields
\bea  
I_{\text{JT cone +GHY}}&=&I_{\text{JT cone}}+I_{GHY}(\text{stretch})=\Delta \log \( \frac{\Lambda_r}{\epsilon_r} \),\\
 Z&=&e^{-I_{\text{JT cone +GHY}}}=\(\frac{\epsilon_r}{\Lambda_r}\)^\Delta.
\eea  
Therefore, the result is the same as the one in higher dimensional cases. On the other hand, in 2D, the SCT reduces to
\bea  
T=\frac{t-b(t^2+z^2)}{1-2b t+b^2(t^2+z^2)},\quad Z=\frac{z}{1-2b t+b^2(t^2+z^2)},
\eea  
which maps the wedge $|T|=r_h Z$ to a 2d banana-shape wedge defined by
\be 
\left(t-\frac{1}{2b} \right)^2+ \left(z\pm\frac{r_h^2}{2b} \right)^2=\frac{1}{4b^2}(1+r_h^2).
\ee 
Therefore, indeed we can construct a 2D spacetime banana and its on-shell action will also equal to
\be 
Z=e^{-I_{\text{JT banana}}}=\(\frac{\epsilon}{1/b}\)^{2\Delta}=\langle\mathcal{O}(0)\mathcal{O}(b^{-1})\rangle.
\ee 
This result is both reasonable and expected, as we can interpret JT gravity as the effective theory of the near-extremal BTZ black hole. Notably, JT gravity also possesses a trumpet solution and various (off-shell) wormhole solutions, apart from the black hole solution. The Euclidean path integral over these wormhole solutions corresponds to the correlation functions in certain random matrix theories.  It would be interesting to understand the meaning of the possible wormhole solutions that connect several spacetime bananas.
\subsection{Holographic local quench}
In \cite{Nozaki:2013wia}, the authors proposed a back-reacted geometry for a falling massive particle, interpreted as a holographic model of local quench. The geometry is obtained by applying the following one-parameter map
\bea 
&&\sqrt{1+r^2}\cos\tau=\frac{\alpha^2+(z^2+x^2+t^2)}{2\alpha z},\nonumber \\
&&\sqrt{1+r^2}\sin\tau=\frac{t}{z},\nonumber\\
&&r\sin\theta=\frac{x}{z},\nonumber \\
&&-r\cos\theta=\frac{-\alpha^2+z^2+x^2+t^2}{2\alpha z}, \label{hlq}
\eea 
to the conical AdS geometry or BTZ black hole. Later on in \cite{Kawamoto:2022etl}, it was generalized to AdS/BCFT. We find that this one-parameter map is a special combination of GtP, SCT and translation. Therefore, this holographic local quench geometry can be viewed as a special case of the spacetime banana. The authors considered the Lorentzian signature of this geometry and by computing the holographic entanglement entropy they found that the dual CFT state is pure regardless of whether it is constructed from a conical AdS or BTZ black hole. This result is very puzzling to us. Because in the cone or banana coordinates, the horizon of the black hole still exists, we expect the geometry will still exhibit some thermal properties. For example, the RT surface may have a phase transition like in a black hole geometry. In the Euclidean spacetime banana, for computing the correlation function, we also put a hard cut-off surface at the stretched horizon which plays the role of projecting the canonical ensemble to a pure state. It should be pointed out that in Lorentzian geometry, the meaning of such a cut-off surface is not clear. We leave these interesting questions for future investigation. 
\section*{Acknowledgments}
We thank Jun Nian, Cheng Peng, Jie-qiang Wu and Ding-fang Zeng for the valuable discussions. We are also very grateful to the anonymous referee for the very helpful remarks regarding the tensor structure of the correlation functions. JT is supported by the National Youth Fund No.12105289 and funds from the University of Chinese Academy of Sciences (UCAS), the Kavli Institute for Theoretical Sciences (KITS). FO would like to thank Masahiro Nozaki very much for his support. The work of FO was supported by funds from UCAS and KITS.
\appendix
\section{Holographic correlation functions of the twist operators}
\label{appendix:twist}
\renewcommand{\theequation}{A.\arabic{equation}}
\setcounter{equation}{0}
In this appendix, we review the holographic calculation of the two-point function of the twist operators following \cite{Hung:2011nu}. The action considered there is the standard Euclidean AdS Hilbert-Einstein action
\be 
I_{}=-\frac{1}{16\pi G_N}\int d^3x\sqrt{g}(\mathcal{R}+2)-\frac{1}{8\pi G_N}\int_{\partial_\text{AdS}}\sqrt{h}(\mathcal{K}-1) \label{EAdS}.
\ee 
The twist operator is a heavy operator with the conformal dimension 
\be 
h=\bar{h}=\frac{c}{24} \( q-\frac{1}{q} \),\quad \Delta=h+\bar{h}=\frac{c}{12} \( q-\frac{1}{q} \), \label{deltat}
\ee 
where $q>1$ is the index of the twist operator. Using the replica trick, we can relate the two-point function of the twist operators to the CFT partition functions via
\be 
\langle \sigma_q(v_1)\bar{\sigma}_q(v_2)\rangle=\frac{Z_q}{Z_1^q},
\ee 
where $Z_q(Z_1)$  is the partition function of the replicated manifold (original manifold). The replicated manifold (after a quotient) has two singular points at the insertion points $v_1$ and $v_2$ where there is an angular excess of $2\pi(q-1)$. On this singular manifold, we can introduce a metric
\be 
ds^2_{Z_q}=\frac{(v_2-v_1)^2}{q^2}\frac{du d\bar{u}}{|(u-v_1)^{(1-1/q)}(u-v_2)^{(1+1/q)}|^2}.
\ee 
Then, by applying the following conformal transformation
\be 
U=\(\frac{u-v_1}{u-v_2}\)^{\frac{1}{q}},
\ee 
the metric can be transformed into a smooth one
\be
ds_{Z_1}^2=dUd\bar{U}, \label{smooth}
\ee 
which can be identified with the original manifold.
To compute $Z_1$, we consider the vacuum bulk solution
\be 
ds^2=\frac{dZ^2+dUd\bar{U}}{Z^2},
\ee 
which is \Poincare AdS$_3$ whose on-shell action vanishes so that $Z_1=0$. To compute $Z_q$, we can use the \Banados bulk solution
\be 
ds^2=\frac{dz^2+dud\bar{u}}{z^2}+L du^2+\bar{L}d\bar{u}^2+ z^2L\bar{L}dud\bar{u}, \label{twistmetric}
\ee 
where 
\be 
L=\frac{3(U'')^2-2U'U'''}{4U'^2}=\frac{(1-q^2)}{4q^2}\frac{(v_1-v_2)^2}{(u-v_1)^2(u-v_2)^2}.
\ee 
Substituting the \Banados solution into the Euclidean AdS Einstein-Hilbert action, one can get
\bea  
I&=&-\frac{c}{6\pi}\int \sqrt{g^{(0)}}\sqrt{L\bar{L}}\label{onfg}\\
&=&-\frac{c}{24\pi} \( \frac{q^2-1}{q^2} \) \int \sqrt{g^{(0)}} \frac{|v_1-v_2|^2}{|u-v_1|^2|u-v_2|^2},\label{twistinteg}
\eea 
where the Brown-Henneaux relation $\frac{1}{2G_N}=\frac{c}{3}$ \cite{Brown:1986nw} has been used.
In \cite{Hung:2011nu}, this integral was computed by observing that the integral \eqref{twistinteg} is a total derivative
\be 
I=-\frac{c}{96\pi}\(1-\frac{1}{q^2}\)\int dud\bar{u}\(\sqrt{g^{(0)}}\nabla_a(\chi g_{(0)}^{ab} \nabla_b\chi)\),\quad \chi=\log|\frac{u-v_1}{u-v^2}|^2,
\ee 
and then using the divergence theorem to rewrite this surface integral as two loop integrals around each singular point
\bea  
I&=&-\frac{c}{96\pi}\(1-\frac{1}{q^2}\)\(\oint_{v_1}+\oint_{v_2}\)d\theta \, r \chi\partial_r\chi \label{divergence}\\
&=&-\frac{c}{96\pi}\(1-\frac{1}{q^2}\)\(\oint_{v_1}+\oint_{v_2}\)d\theta \, 4\log \( \frac{\epsilon_r}{|v_1-v_2|} \) \nn
&=&\frac{c}{6}\(q-\frac{1}{q}\)\log \( \frac{|v_1-v_2|}{\epsilon_r} \) 
\nn
&=& \Delta \log \( \frac{|v_1-v_2|}{\epsilon_r} \). \label{hung}
\eea  
However, one can also directly compute the integral \eqref{twistinteg} by introducing a Feynman parameter then regulating the resulting integral by the dimensional regularization. The result is 
\bea  
I&=&-\frac{c}{24\pi}\(1-\frac{1}{q^2}\)\frac{|v_1-v_2|^22\pi q}{|v_{1}-v_2|^2}\(\frac{2}{\epsilon_d}+2\log|v_1-v_2|+\gamma+\log\pi\)\\
&\sim& -\frac{c}{6}\(q-\frac{1}{q}\)\log|v_1-v_2|
\nn
&=& -\Delta \log|v_1-v_2|, \label{dimenr}
\eea  
which differs from \eqref{hung} by a minus sign. We believe that the second result  \eqref{dimenr} is correct. The reason for this is as follows: when we remove the two small disks centered at each singular point, we create two circular boundaries on the manifold. It is important to note that the normal vector of these circular boundaries should point inward which is toward the center. Therefore, when we use the divergence theorem to derive equation \eqref{divergence}, we need to include an extra minus sign. In other words, the Euclidean AdS Einstein-Hilbert partition function of the \Banados solution \eqref{twistmetric} does \textit{not} equal to the two-point function but the inverse of it. Below we give another explanation in support of this conclusion.
\subsection{One-point function}
One can try a similar idea to compute the one-point function of the twist operator. If we put one twist operator in the origin, we expect that
\be 
\langle \sigma_q\rangle=Z_q/Z_1 ,
\ee 
where $Z_q$ is related to a manifold with a conical singularity of angle $2\pi(q-1)$. On the manifold, we can introduce the following metric
\be 
ds^2_{Z_q}=\frac{1}{q^2}|u|^{2(\frac{1}{q}-1)}dud\bar{u}.
\ee 
Similarly, we can apply the conformal transformation
\be 
U=u^{\frac{1}{q}},
\ee 
to get the smooth metric \eqref{smooth}. The \Banados solution used in the calculation of $Z_q$ is described by \eqref{twistmetric} with
\be 
L=-\frac{(q^2-1)}{4q^2}\frac{1}{u^2}. \label{onetwist}
\ee 
Therefore, the on-shell action is given by
\be 
I=-\frac{c}{6\pi}\frac{(q^2-1)}{4q^2}2\pi q \int_{\epsilon_r}^{\Lambda_r} dr \frac{r}{r^2}=-\Delta \log \( \frac{\Lambda_r}{\epsilon_r} \) ,\label{onshell_onetwist}
\ee 
and hence the partition function is as follows
\be 
Z_q=e^{-I}=\(\frac{\Lambda_r}{\epsilon_r}\)^{\Delta }.
\ee 
Notice that it is the inverse of the one-point function on a disk with radius $\Lambda_r$:
\be 
\langle \sigma_q(0,0)\rangle_{\text{disk}}=\(\frac{\Lambda_r}{\epsilon_r}\)^{-\Delta }.
\ee 
\section{Conical geometry}
\label{appendix:particle}
\renewcommand{\theequation}{B.\arabic{equation}}
\setcounter{equation}{0}
In this appendix, we study the AdS conical geometry in various coordinates. 
\subsection{One-point function}
In the global coordinates, the AdS$_3$ conical geometry has the following metric
\be \label{conicalmetric}
ds^2=(r^2+a^2)d\tau^2+\frac{dr^2}{(r^2+a^2)}+r^2d\theta^2,
\ee  
where the parameter $a$ is related to the mass of the particle at the center via $4 G_Nm=1-a>0$. Assuming the time direction is compact $\tau\sim \tau+\beta$, we find that the corresponding on-shell action is given by
\be 
I_{\text{AdS-EH}}=-\frac{a^2 \beta}{8G_N} \equiv \( \Delta-\frac{c}{12} \)\beta, \label{mdelta}
\ee 
which implies that the relation between the mass of the particle and the conformal dimension of the dual operator is as follows
\be 
m=\frac{c}{6}\(1-\sqrt{1-\frac{12\Delta}{c}}\).
\ee 
Starting from it, we first use the GtP map
\be
r=\frac{R}{z},\quad \tau=\frac{1}{2}\log\(R^2+z^2\),
\ee
to get the deformed \Poincare spacetime with the following metric
\bea\label{ccone}
ds^2=\frac{1}{z^2}\[\frac{\(z dR-R dz\)^2}{R^2+a^2z^2}+\frac{\(R^2+a^2z^2\)\(R dR+z dz\)^2}{\(R^2+z^2\)^2}+R^2d\q^2\].
\eea
Clearly, the position of the conical singularity is mapped to 
\be
R=0,\quad z=e^{\tau},
\ee
so the solution \eqref{ccone} also describes a back-reacted geometry with one heavy operator inserted at the origin. It is straightforward to find
\be
\sqrt{g}=\frac{R}{z^3},\quad \sqrt{h\big|_{\pa}}=\frac{R}{\e^2}-\frac{1-a^2}{2R}+\mathcal{O}\(\e^2\),\quad \mathcal{K}\big|_{\pa}=2+\mathcal{O}(\e^4),
\ee
and the on-shell action is equal to
\bea
I_{\text{AdS-EH}}&=&-\frac{1}{16\pi G_N}\int d^{3}x\sqrt{g}\(\mathcal{R}-2\L\)-\frac{1}{8\pi G_N}\int d^{2}x\sqrt{h}\(\mathcal{K}-1\)\\
&=&\frac{1}{4\pi G_N}\int_{0}^{2\pi}d\q\int_{0}^{\infty} R dR\int_{\e}^{\infty}\frac{dz}{z^3}-\frac{1}{8\pi G_N}\int_{0}^{2\pi} d \theta \int_{0}^{\infty}dR\(\frac{R}{\e^2}-\frac{1-a^2}{2R}\)\nonumber\\
&=&\frac{(1-a^2)}{8G_N}\int_{\e_0}^{\L_0}\frac{dR}{R}=\frac{(1-a^2)}{8G_N}\log\frac{\L_0}{\e_0}=\Delta \log \( \frac{\L_0}{\e_0} \), \label{onshellccone}
\eea
where we have used \eqref{mdelta} and introduced radial cut-offs $\epsilon_0$ and $\Lambda_0$. As promised, the partition function takes the form of one-point correlation function
\be 
e^{-I_{\text{AdS-EH}}}=\(\frac{\epsilon_0}{\Lambda_0}\)^\Delta.
\ee 
Next, we compute the same on-shell action in the FG coordinates $(R_{f},z_{f},\theta_{f})$ which are related to \eqref{ccone} via
\be\label{FG coordinate-transformation}
\frac{R}{z}=\frac{R_{f}}{z_{f}}\(1+\frac{(1-a^2)}{4}\frac{z_{f}^2}{R_{f}^2}\),\quad R^2+z^2=R_{f}^2\(1+\frac{4a z_{f}^2}{4R_{f}^2+\(1-a\)^2z_{f}^2}\)^{\frac{1}{a}}.
\ee
%it's direct to solve the FG expansion of the above metric
%\be\label{one-point-FG}
%ds^2=\frac{dz^2}{z^2}+\frac{\(4R^2-(1-a^2)z^2\)^2dR^2+\(4R^2+(1-a^2)z^2\)^2R^2d\q^2}{16R^4z^2},
%\ee
Introducing the light-cone coordinates
\be
u_{f}=R_{f} e^{\text{i}\theta_{f}},\quad \bar{u}_{f}=R_{f} e^{-\text{i}\theta_{f}},
\ee
we can rewrite the FG gauge in the standard form of a \Banados solution
\be\label{one-point-LC-FG}
ds^2=\frac{dz_{f}^2+\(du_{f}+z_{f}^2\Bar{L}(\bar{u}_{f})d\bar{u}_{f}\)\(d\bar{u}_{f}+z_{f}^2L(u_{f})du_{f}\)}{z_{f}^2},
\ee
with
\be
L(u_{f})=-\frac{1-a^2}{4u_{f}^2},\quad \Bar{L}(\bar{u}_{f})=-\frac{1-a^2}{4\bar{u}_{f}^2}.\label{oneconical}
\ee
This \Banados solution has a coordinate singularity at the position
\be
\frac{R_{f}}{z_{f}}=\frac{\sqrt{1-a^2}}{2},
\ee
where the determinant of the metric
\be
\sqrt{g}=\frac{16R_{f}^4-\(1-a^2\)^2z_{f}^4}{16R_{f}^3z_{f}^3},
\ee
vanishes. This coordinate singularity is referred to as the wall of the \Banados solution. It should be pointed out that the FG coordinates only covers the region outside the wall \ie $R_{f}/z_{f}\geq \sqrt{1-a^2}/2$, so the on-shell action is
\bea
I_{FG}&=&\frac{1}{4\pi G_N}\int_{0}^{2\pi}d\q\int_{\e_0}^{\L_0}dR_{f}\int_{\e}^{\frac{2R_{f}}{\sqrt{1-a^2}}}\(\frac{R_{f}}{z_{f}^3}-\frac{\(1-a^2\)^2}{16R_{f}^3}z_{f}\)dz_{f}
\nonumber \\
&& -\frac{1}{8\pi G_N}\int_{0}^{2\pi}d\q\int_{\e_0}^{\L_0}\frac{R_{f}}{\e^2}dR_{f}\nonumber\\
&=&\frac{1}{2G_N}\left.\[-\frac{(1-a^2)}{4}\log R_{f}+\frac{R_{f}^2}{4\e^2}\]\right|_{\e_0}^{\L_0}-\left.\frac{R_{f}^2}{8\pi G_N\e^2}\right|_{\e_0}^{\L_0}
\nonumber \\
&=& -\frac{(1-a^2)}{8G_N}\log \( \frac{\L_0}{\e_0} \),
\eea
which differs from \eqref{onshellccone} by a minus sign! The discrepancy can be understood as follows. Using the relations \eqref{FG coordinate-transformation}, we find that the condition $R_{f}/z_{f}\geq \sqrt{1-a^2}/2$ is equivalent to
\be 
\frac{R}{z}\geq \sqrt{1-a^2}, \label{misscone}
\ee 
which implies that the FG gauge misses the region inside the cone defined by \eqref{misscone}. Furthermore, we find that the on-shell action inside the cone is\footnote{here we have set $\e=\e_0/\sqrt{1-a^2}$ in the final step.}
\bea
I_{cone}&=&\frac{1}{2 G_N}\[\int_{\e_0}^{\sqrt{1-a^2}\e}R dR\int_{\e}^{\infty}\frac{dz}{z^3}+\int_{\sqrt{1-a^2}\e}^{\infty}R dR\int_{\frac{R}{\sqrt{1-a^2}}}^{\infty}\frac{dz}{z^3} \right.\nn
&\,&\left. -\frac{1}{2}\int_{\e_0}^{\sqrt{1-a^2}\e}\(\frac{R}{\e^2}-\frac{1-a^2}{2R}\)dR\]\nonumber\\
&=&\frac{(1-a^2)}{4G_N}\[\log \( \frac{\L_0}{\sqrt{1-a^2}\e} \) +\frac{1}{2}\log \( \frac{\sqrt{1-a^2}\e}{\e_0} \) \]
\nonumber\\
&=& \frac{(1-a^2)}{4G_N}\log \( \frac{\L_0}{\e_0} \),
\eea
which leads to a reasonable result
\be
I_{FG}+I_{cone}=I_{\text{AdS-EH}}. \label{nice}
\ee
To address the sign error in the on-shell action within the FG gauge, one might consider adding a GHY term either at the wall in the FG gauge or at the boundary of the cone in global coordinates. However, explicit calculations show that these modifications do not work.
\subsection{Two-point function}
To introduce an additional insertion point, the next step involves performing the SCT transformation \eqref{stp} (with $b_t=0$).
The resulting metric is complicated and not particularly insightful, so we omit it here. However, the determinants of both the metric and induced metric at the AdS boundary are quite straightforward \footnote{Let the coordinates be denoted as  $(z,t,x)$; we hope this does not cause any confusion.}
\be 
\sqrt{g}=\frac{1}{z^3},\quad \sqrt{h}=\frac{1}{\epsilon^2}-\frac{1-a^2}{2(t^2+x^2)(b_x^2t^2+(b_xx-1)^2)},\quad \mathcal{K}=2.
\ee 
The on-shell action is 
\bea  
I_{\text{AdS-EH}}&=&\frac{(1-a^2)}{16\pi G_N}\int dtdx\frac{1}{(t^2+x^2)(b_x^2t^2+(b_xx-1)^2)} \label{integral}\\
&=&\frac{(1-a^2)}{8G_N}\log\left|\frac{1/b_x}{\epsilon}\right|^2=\Delta \log\left|\frac{1/b_x}{\epsilon}\right|^2\equiv\Delta \log\left|\frac{x_{12}}{\epsilon}\right|^2 ,
\eea  
where we used the fact that the two inserted operators are located at $\vec{x}_1=(0,0)$ and $\vec{x}_2=(0,1/b_x)$. The integral \eqref{integral} is equivalent to \eqref{twistinteg} and can be evaluated using various techniques. The integrand contains two poles, indicating that the integral is naively divergent and requires regularization. Consequently, the finite part of the result depends on the regularization scheme, while the leading divergent part should be consistent across all methods. Here, we present a ``brute force" approach. We find that the integral can be computed by decomposing the integral domain as follows:
\be 
\int_{-\infty}^\infty dt\int_{-\infty}^\infty dx=\int_{-\infty}^\infty dt\int_{-\infty}^{0-\epsilon}dx+\int_{-\infty}^\infty dt\int_{\frac{1}{b_x}+\epsilon}^{\infty}dx+\int_{-\infty}^{0-\epsilon}dt\int_{0}^{\frac{1}{b_x}}dx+\int^{\infty}_{0+\epsilon}dt\int_{0}^{\frac{1}{b_x}}dx.
\ee 
In each subdomain, the integral remains finite, and the leading result in $\epsilon$ is $-\pi\log(b_x\epsilon) $. Indeed, the partition function reproduces the two-point function
\be 
e^{-I_{\text{AdS-EH}}}=\(\frac{\epsilon}{|x_{12}|}\)^{2\Delta}.
\ee 
Next, let us compute the on-shell action in the FG gauge which can be found by imposing the following ansatz 
\be 
t\rightarrow t+c_4 z^4+c_6 z^6+\dots,\quad x\rightarrow x+d_4 z^4 +d_6z^6+\dots,\quad z\rightarrow z+f_3 z^3+f_5 z^5+\dots.
\ee 
The resulting metric takes the standard form of the \Banados solution \eqref{one-point-LC-FG}, with 
\be 
L(u)=-\frac{1-a^2}{4u^2(b_x u-1)^2},\quad \bar{L}(v)=-\frac{1-a^2}{4v^2(b_x v-1)^2}.
\ee 
Substituting them into \eqref{twistinteg}, we get
\bea  
I_{FG}&=&-\frac{c}{6\pi}\frac{(1-a^2)}{4}\int dt dx\frac{1}{(t^2+x^2)(b_x^2t^2+(b_xx-1)^2)}\nn
&=&-\frac{(1-a^2)}{16\pi G_N}\int dtdx\frac{1}{(t^2+x^2)(b_x^2t^2+(b_xx-1)^2)},
\eea  
which again differs from \eqref{integral} by a minus sign. Note that the metric in the FG coordinates also has a  wall at the position which is determined by the following equation
\be 
z^2=\frac{1}{\sqrt{L(u)\bar{L}(v)}}=\frac{2}{\sqrt{1-a^2}}{(t^2+x^2)(b_x^2t^2+(b_xx-1)^2)}.
\ee 
Unfortunately, in this case we do not have a closed form of the transformation between the FG gauge and the original coordinates such that we can obtain a nice result similar to \eqref{nice}. 
%%%%%%%%%%%%%%%%%%%%%%%%%%
\section{Revisit the world-sheet banana}
%%%%%%%%%%%%%%%%%%%%%%%%%%
\label{appendix:string}
\renewcommand{\theequation}{C.\arabic{equation}}
\setcounter{equation}{0}
In this appendix, we revisit a closely related idea \cite{Janik:2010gc} concerning the computation of correlation functions for local operators associated with classical spinning string states. Notably, the string solution resembles a banana which can be referred to as the world-sheet banana. Strings are not simple local objects, and hence they have non-trivial internal degrees of freedom. The Euclidean partition function of string theory receives contributions from an infinite number of string states. To derive the correlation functions for specific states, it is necessary to modify the naive classical action through a projection onto the appropriate wave functions. The same approach has been adopted in the context of the spacetime banana, where the projection is implemented by introducing a stretched horizon or an EOW  brane. For strings, in the work by Janik et al. \cite{Janik:2010gc}, it was proposed that the correlation function can be computed by first evaluating the Lorentzian Polyakov action for classical string solution, followed by a Legendre transformation that converts the Lagrangian into the Hamiltonian, then extremizing the result with respect to the modular parameter of the string world-sheet. For a detailed discussion of this approach, we refer the readers to the original work \cite{Janik:2010gc}. The remainder of this appendix is dedicated to the study of the rotating string solution in AdS$_3$.
\\We start from the equation of motion of the probe string. It is much simpler to write them in the embedding coordinates. Let us parameterize EAdS$_d$ with $d+1$ embedding coordinates $\vec{Y}$ subject to the following constraint
\be 
\vec{Y}\cdot \vec{Y}=-1.
\ee 
In the conformal gauge $z=\frac{1}{2}(\sigma-\tau)$, the equation of motion and the Virasoro constraints are as follows
\bea 
&& \partial \bar{\partial}\vec{Y}-(\partial \vec{Y}\cdot \bar{\partial}\vec{Y})\vec{Y}=0,\\
&& \partial \vec{Y}\cdot \partial \vec{Y}=\bar{\partial} \vec{Y}\cdot \bar{\partial} \vec{Y}=0.
\eea 
The spinning string solution has the following form in the global coordinates on EAdS$_3$:
\bea 
&&Y_0=\cosh\rho_0 \cosh  k\tau,\quad Y_1=\cosh\rho_0 \sinh k\tau,\\
&&Y_2= \sinh\rho_0\cos(\w \tau+\sigma),\quad Y_3= \sinh\rho_0\sin(\w \tau+\sigma),\\
&&-Y_0^2+Y_1^2+Y_2^2+Y_3^3=-1.
\eea
The equation of motion imposes the following condition
\be 
k^2=1-\w^2,\label{eqm}
\ee 
but the Virasoro constraint has no solutions. To satisfy the Virasoro constraints, we need the string also rotate in some internal spacetime, say $S^1$:
\bea 
X_1=\cos(\nu\tau-\sigma),\quad X_2=\sin(\nu\tau-\sigma),
\eea 
then the two Virasoro constraints are as follows
\bea  
1+k^2+\nu^2+(1+k^2+\omega^2)\sinh^2(\rho_0)&=&0,\label{v1}\\
\nu-\omega \sinh^2(\rho_0)&=&0\, .
\eea   
On the other hand, the angular momenta of this string solution are given by \footnote{To compare with the results in \cite{Janik:2010gc}, here we also omit the factor $\sqrt{\lambda}$.} 
\bea 
J_{01}&=&\frac{1}{2\pi}\int_0^{2\pi}d\sigma (Y_0\dot{Y}_1-Y_1\dot{Y}_0)=k\cosh^2(\rho_0),\label{c1}\\
J_{23}&=&\frac{1}{2\pi}\int_0^{2\pi}d\sigma (Y_2\dot{Y}_3-Y_2\dot{Y}_3)=\omega \sinh^2(\rho_0),\label{c2}\\
S_{12}&=&\frac{1}{2\pi}\int_0^{2\pi}d\sigma (X_1\dot{Y}_2-X_2\dot{X}_1)=\nu.
\eea 
Moreover, the modified Polyakov action is \cite{Janik:2010gc}  
\bea  
S_p&=&-\frac{\sqrt{\lambda}}{4\pi}\int_{0}^{s}d\tau\int_{0}^{2\pi}d\sigma\(-\dot{\vec{V}}\cdot \dot{\vec{V}}+ {\vec{V}'}\cdot {\vec{V}'}\)\nn 
&~&-\int_{0}^{s}d\tau\int_{0}^{2\pi}d\sigma(\vec{\Pi}-\vec{\Pi}_0)\cdot (\dot{\vec{V}}-\dot{\vec{V}}_0), \label{Polyakov}
\eea
where $s$ is the world-sheet modular parameter, $\vec{V}$ denotes the collective embedding coordinates $(\vec{Y},\vec{X})$, $\vec{\Pi}$ is the conjugate momentum and $\vec{\Pi}_0,\dot{\vec{V}}_0$ are the zero modes defined as follows
\be 
\vec{\Pi}_0=\frac{1}{2\pi}\int_0^{2\pi}d\sigma \vec{\Pi},\quad \dot{\vec{V}}_0=\frac{1}{2\pi}\int_0^{2\pi}d\sigma \dot{\vec{V}}.
\ee 
Then, by substituting the spinning string solution into \eqref{Polyakov}, we get
\be 
S_p=\frac{\sqrt{\lambda}}{2}\(k^2+2\sinh^2\rho_0(k^2-1)-(1+\nu^2)\)s.
\ee 
To extremize this action, we first need to impose the proper boundary conditions. In the global EAdS$_3$ coordinates, the string rotates in the angular direction at a fixed radius $\rho=\rho_0$ with an angular velocity $\omega$ and propagates from $\tau=0$ to $\tau=\tau_\infty$ with a velocity $k$. For convenience, we can introduce a  cut-off for the Euclidean time $\tau_\infty=\beta$, thus in the string world-sheet, we should impose $ ks=\beta$, which means that extremizing  the action with respect to $s$ is equivalent to extremizing with respect to $k$.
Because we want to pick out a specific string state, we also fix the other two charges $J_{23}$ and $S_{12}$ of the string solution. Therefore, we should rewrite the action as follows
\be 
S_p=\beta\frac{\sqrt{\lambda}}{2}\frac{\( k^2-1-2J_{23}\sqrt{1-k^2}-S_{12}^2 \)}{2k},\label{sfix}
\ee 
then the saddle point approximation with respect to $k$ leads to
\be
1+k^2+\frac{2J_{23}}{\sqrt{1-k^2}}+S_{12}^2=0. \label{saddle}
\ee 
Solving $S_{12}$ from the saddle point equation and substituting the solution into \eqref{sfix}, we get
\be 
S_p=\beta k\sqrt{\lambda}\(1+\frac{J_{23}}{\sqrt{1-k^2}}\)=\beta \sqrt{\lambda}J_{01}\equiv \beta E,
\ee 
where in the last equality we have used \eqref{eqm},\eqref{c1} and \eqref{c2}. Moreover, in the last equality we have identified  the boost charge $\sqrt{\lambda}J_{01}$ as the string energy $E$ or the conformal dimension of the string state.
\\Note that this rotating string solution always stays at a fixed radial position. Therefore, we can apply the GtP map to the string solution to get a cone solution. Recalling that the relations between the embedding coordinates and the \Poincare coordinates are as follows
\bea 
&&Y_0=\frac{1+Z^2+X^2+T^2}{2Z},\quad Y_1=\frac{-1+Z^2+X^2+T^2}{2Z},\nn
&&Y_2=\frac{T}{Z},\quad Y_3=\frac{X}{Z},
\eea 
we can directly find that the string world-sheet profile in the \Poincare AdS spacetime is 
\be 
T=\tanh\rho_0 \cos(\omega \tau+\sigma)e^{k\tau},\quad X=\tanh\rho_0 \sin(\omega \tau+\sigma)e^{k\tau},\quad Z=\frac{1}{\cosh\rho_0}e^{\kappa \tau}.
\ee 
Then, by introducing the new coordinate $R$ as follows
\bea 
T^2+X^2=R^2,
\eea 
we find that the string world-sheet indeed sweeps a cone:
\bea 
R=\tanh\rho_0 e^{\kappa \tau},\quad Z=\frac{1}{\cosh\rho_0}e^{\kappa \tau},\quad \frac{Z}{R}=\frac{1}{\sinh\rho_0}.
\eea 
We would like the string world-sheet to cover the whole region of the cone. Thus, we choose the following boundary condition
\be 
\int_{s_i}^{s_f} d\tau=\frac{1}{k}\int_{\epsilon}^{\Lambda} d[\log(\cosh\rho_0 Z)]=\frac{1}{k}\log \( \frac{\Lambda}{\epsilon} \).
\ee  
Therefore, the modified Polyakov action becomes
\bea  
S_p&=&\frac{\sqrt{\lambda}}{2}\(k^2+2\sinh^2\rho_0(k^2-1)-(1+\nu^2)\)s \nn
&=&\frac{\sqrt{\lambda}}{2}\frac{\left( k^2-1-2J_{23}\sqrt{1-k^2}-S_{12}^2 \right)}{2k}\log \( \frac{\Lambda}{\epsilon} \).
\eea  
The saddle point equation with respect to $k$ is still \eqref{saddle}, so at the saddle point the action is equal to
\be 
S_p=E\log \( \frac{\Lambda}{\epsilon} \).
\ee  
Therefore, we reproduce the one-point function
\be 
e^{-S_p}=\(\frac{\epsilon}{\Lambda}\)^{\Delta},\quad \Delta=E.
\ee 
Next, we can apply the boost \eqref{stp} on the cone world-sheet to get the banana world-sheet. Let us focus on a special section of the world-sheet profile in the $Z$ component
\bea  
z&=&\frac{1}{e^{-k \tau}\cosh\rho_0+(b_t^2+b_x^2)e^{k\tau}\cosh\rho_0+2\sinh\rho_0(b_t\cos(\sigma+\omega \tau)+b_x \sin(\sigma+\omega \tau))}\nn 
&\sim &\frac{1}{e^{-k \tau}\cosh\rho_0+(b_t^2+b_x^2)e^{k\tau}\cosh\rho_0}. \label{stringsection}
\eea  
When $\tau=\pm \infty$, we have $z\rightarrow 0$, so the world-sheet touches the AdS boundary at two points as desired. The maximal value of $z$ is given by
\be 
z_{max}=\frac{|\vec{b}|}{2 \cosh\rho_0}.
\ee 
Therefore, we should impose the following boundary condition
\be 
\int_{s_i}^{s_f} d\tau=2\int_\epsilon^{z_{max}}d[\tau(z)]=\frac{2}{k}\log \( \frac{|b|}{\epsilon \cosh\rho_0} \) \sim \frac{1}{k}\log\(\frac{|b|}{\epsilon}\)^2, 
\ee 
where we have absorbed $\cosh\rho_0$ into $\epsilon$ as suggested in \cite{Janik:2010gc}. Then, by following the same procedure, we get
\be 
S_p=E\log\(\frac{|b|}{\epsilon}\)^2,\quad e^{-S_p}=\(\frac{\epsilon}{|b|}\)^{2\Delta}. \label{actionspin}
\ee 
Recall that two-point function of operators with spin in 2d CFT is given by \cite{Belavin:1984vu}
\be 
\langle \mathcal{S}(0)\mathcal{S}(v)\rangle=\frac{1}{|v|^{2\Delta}}\(\frac{\bar{v}}{v}\)^J, 
\ee 
thus, the modified Polyakov action \eqref{actionspin} only reproduces the modulus of the correlation function 
\be 
e^{-S_p}\sim |\langle \mathcal{S}(0)\mathcal{S}(v)\rangle|\, .\label{modulus}
\ee 
This apparent inconsistency is resolved in \cite{Kazama:2011cp,Kazama:2013qsa}, where the key proposal involves introducing a semiclassical wavefunction for each operator and reformulating these wavefunctions through specific action-angle variables. Consequently, the total effective action combining both the wavefunctions and string worldsheet action can be expressed as
\begin{equation}
S_p = E (s_f - s_i) - \mathrm{i} J \Omega,
\label{total_action}
\end{equation}
where an explicitly imaginary term emerges. Here $\Omega$ denotes the angular variable conjugate to the action variable $J$. As demonstrated in \cite{Kazama:2011cp,Kazama:2013qsa}, this angular parameter $\Omega$ connects to the boost transformation argument through
\begin{equation}
\Omega = -\frac{\mathrm{i}}{2} \ln \left( \frac{\bar{b}^2}{b^2} \right).
\label{angular_relation}
\end{equation}
Incorporating this additional contribution ultimately yields the complete two-point function
\begin{equation}
e^{-S_p} = \left( \frac{\epsilon}{|b|} \right)^{2\Delta} \left( \frac{\bar{b}}{b} \right)^J.
\label{final_result}
\end{equation} 
%%%%%%%%%%%%%%%%
\section{Visualizing the spacetime banana with EOW brane}
\label{appendix:bananaEOW}
\renewcommand{\theequation}{D.\arabic{equation}}
\setcounter{equation}{0}
In this appendix, we first demonstrate the EOW brane \eqref{bananaEOW} indeed has a profile of a banana, then we verify the cancellation of the surface integrals along the EOW brane by zooming in each section. Of course, one can numerically plot the EOW brane profile and find the banana shape. Here, we present a more intuitive and geometric way to view this which also applies to the original spacetime banana. We begin by selecting a constant $\Theta=\Theta_0$ slice of the  cone EOW brane profile
\be 
Z=R\frac{\sinh(r_h(\pi\pm \Theta))}{\lambda r_h}. 
\ee 
Then the image of this slice in the banana geometry can be found by looking at the overlap of the images of $Y=\tan(\Theta_0)X$ and
\be 
Z=R\frac{\sinh(r_h\Theta)}{\lambda r_h}\Big|_{\Theta=\Theta_0}=X\frac{\sinh{r_h(\pi\pm\Theta_0)}}{\lambda r_h\cos\Theta_0}\equiv  c_{\pm}X.
\label{i2}
\ee 
According to the SCT map (with $b_t=0$), the image of \eqref{i2} is a sphere
\be 
t^2+ \left( x-\frac{1}{2b_x} \right)^2+ \left( z+ \frac{1}{2b_x c_{\pm}} \right)^2=\frac{1}{4b_x^2}\(1+\frac{1}{c_{\pm}^2}\),
\ee 
and the image of $Y=\tan(\Theta_0)X$ is also a sphere
\be 
z^2+ \left( x-\frac{1}{2b_x} \right)^2+ \left( t+\frac{1}{2b_x\tan\Theta_0} \right)^2=\frac{1}{4b_x^2}\(1+\frac{1}{\tan^2\Theta_0}\).
\ee  
Therefore, the overlap of these two spheres is a portion of a circle connecting the two fixed points $(z,x,t)=(0,0,0)$ and $(z,x,t)=(0,\frac{1}{b_x},0)$ regardless of the angle $\Theta_0$ and it lies in the plane
\be 
z=\frac{c_{\pm}}{\tan\Theta_0}t=\frac{\sinh(r_h(\pi\pm\Theta_0))}{\lambda r_h\sin(\Theta_0)} t\equiv  k_\pm t.
\ee 
When we vary $\Theta_0$, the curve will sweep a region in the shape of a banana and the width of the banana is determined by  $k_+-k_-$. 
\\As explained in the main text, in the cone coordinates, the cancellation of the surface integrals have already manifested in the integrand. In contrast, in the banana coordinates, the cancellation is more subtle. Therefore, it is very helpful to consider a toy model first.
\subsection{A toy model}
\begin{figure}[h]
\begin{center}
  \includegraphics[scale=0.4]{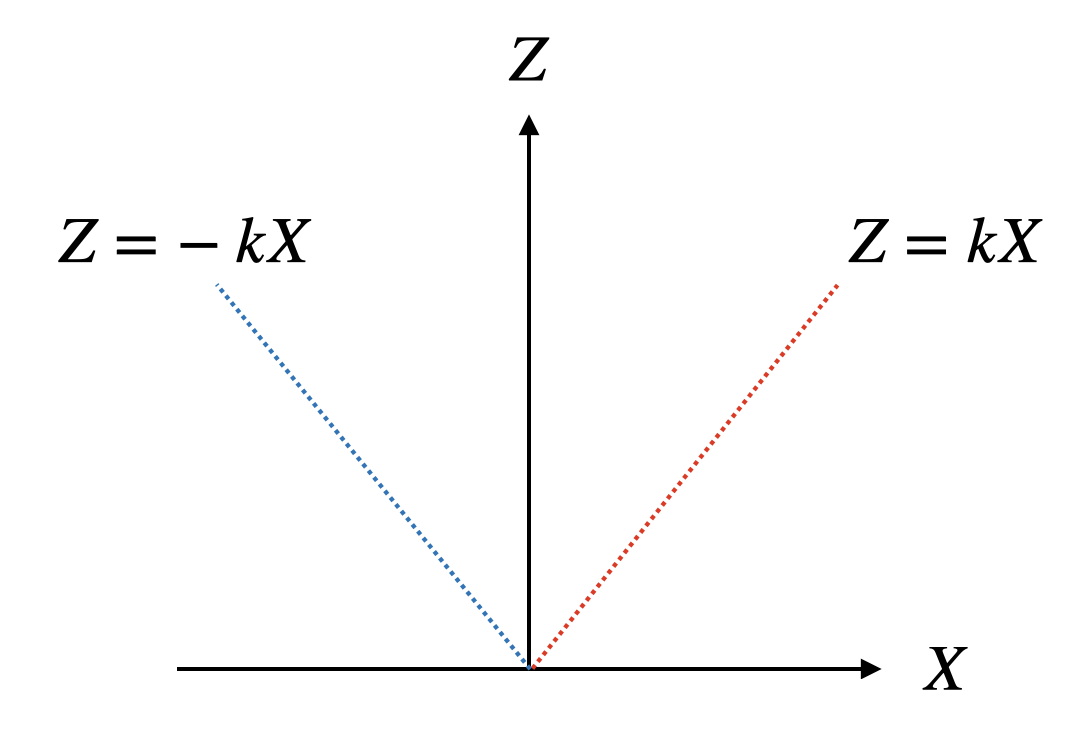}
  \caption{The wedge geometry as a toy model for cone geometry. The dashed blue and red lines are the profile of two planar EOW branes. }\label{wedge}
  \end{center}
\end{figure}

\begin{figure}[h]
\begin{center}
  \includegraphics[scale=0.4]{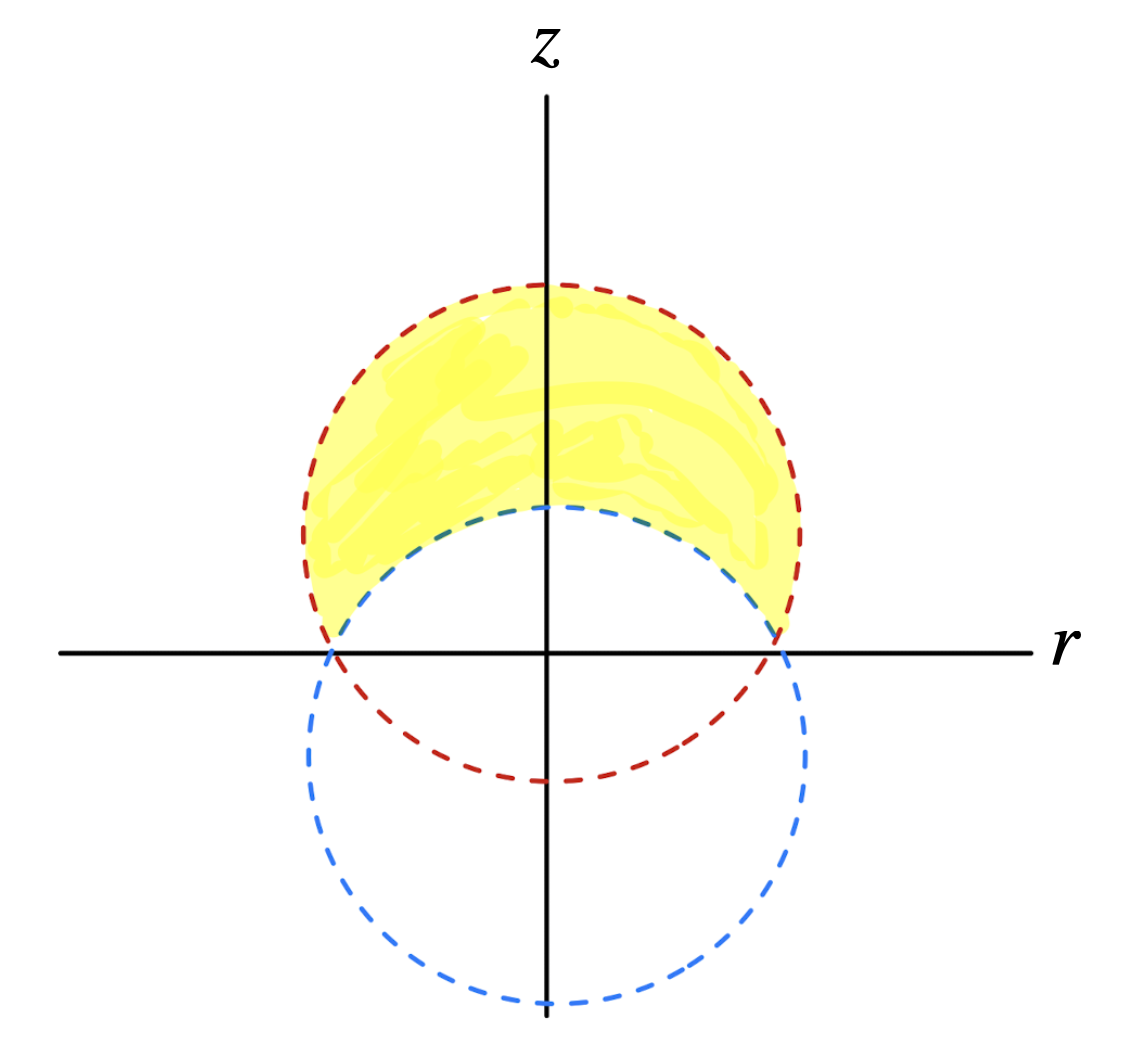}
  \caption{The shape of the wedge geometry in Figure \ref{wedge} after performing the SCT map and a shift in the $x$-direction. Under the SCT map, two planar EOW branes are transformed to spherical branes which are shown by blue and red dashed circles. Moreover, the region inside the cone is mapped to a "bridge-shape" region shown in yellow. The Einstein-Hilbert on-shell action is calculated in this region.}\label{wedge2}
  \end{center}
\end{figure}

Consider the following wedge geometry shown in Figure \eqref{wedge} as a toy model of 
the cone geometry. 
We have two planar EOW branes with the profiles $Z=\pm kX$ in the \Poincare geometry. Since the geometry has a reflection symmetry, we can focus on the right half of it. The on-shell action is given by
\bea  
I_{\text{wedge}}&=&\underbrace{\frac{1}{8\pi G_N}\int dT\int_0^\infty dX\(\frac{1}{\epsilon^2}-\frac{1}{k^2 X^2}\)}_{\text{AdS Einstein-Hilbert}}\underbrace{-\frac{1}{8\pi G_N}\int dT\int_0^\infty dX\frac{1}{\epsilon^2}}_{I_{\text{GHY}}(\partial_{\text{AdS}})+I_{ct}}\nn 
&\,& \underbrace{-\frac{1}{8\pi G_N}\int dT\int_0^\infty dX\frac{\sqrt{1+k^2}}{k^2 X^2}\(-\frac{1}{\sqrt{1+k^2}}\)}_{\text{EOW}},
\eea  
where we see the cancellation happens also at the integrand level. Now we similarly perform the SCT transformations \eqref{stp} (with $b_t=0$), which maps the planar brane in the \Poincare geometry to a spherical brane 
\be 
t^2+ \left( x-\frac{1}{2b_x} \right)^2+ \left( z\pm\frac{1}{2kb_x} \right)^2=\frac{1}{4b_x^2} \left( 1+\frac{1}{k^2} \right)=0,
\ee 
and maps the wedge to a bridge-shape region. For simplicity, we can use the translational symmetry in the $x$-direction to shift the center $x=\frac{1}{2b_x}$ to the origin $x=0$. Figure \eqref{wedge2} demonstrates this configuration.
The bulk part of the on-shell action associated with this region can be reduced to the following surface integrals
\be 
I_{\text{bridge}}=\frac{1}{4G_N}\int_0^{\frac{1}{2b_x}} rdr\(\frac{1}{z_-^2}-\frac{1}{z_+^2}\)+\frac{1}{4G_N}\int_{\frac{1}{2b_x}}^{\frac{\sqrt{1+k^2}}{2b_x k}}rdr\(\frac{1}{z_{+,-}^2}-\frac{1}{z_{+,+}^2}\), \label{bridge}
%\int_{\frac{1}{2b_x}}^{\sqrt{\frac{1+k^2}{4b_x^2k^2}-(z-\frac{1}{2kb_x})^2}}rdr\int_{\epsilon}^{\frac{1}{kb_x}} \frac{1}{z^3}dz 
\ee 
where
\be 
z_\pm=\sqrt{\frac{1+k^2}{4b_x^2k^2}-r^2}\pm\frac{1}{2kb_x},\quad z_{+,\pm}=\frac{1}{2b_xk}\pm \sqrt{\frac{1+k^2}{4b_x^2k^2}-r^2}.
\ee 
Let us examine the surface integrals along the EOW brane described by $z_-$. The EOW brane action is 
\be 
I_{\text{EOW},z_-}=\frac{1}{4G_N}\int_0^{\frac{1}{2b_x}} rdr \frac{1}{z_-^2}\frac{1}{\sqrt{1+k^2(1-4b_x^2r^2)}}.
\ee 
It is clear that the EOW brane action cannot cancel the one presented in equation \eqref{bridge}. In this scenario, the cancellation occurs between the two EOW branes. The action in the bridge region is given by the difference between the actions of the upper and lower hemispheres. Using the result \cite{Fujita:2011fp} of the on-shell action of a hemisphere in AdS/BCFT, we find
\be \label{bridgec}
I_{\text{bridge}}=\underbrace{\frac{1}{4G}\(\log\frac{\epsilon}{1/(2b_x)}-\sinh^{-1}k\)}_{\text{upper hemisphere }}-\underbrace{\frac{1}{4G}\(\log\frac{\epsilon}{1/(2b_x)}-\sinh^{-1}k\)}_{\text{lower hemisphere}}=0.
\ee 
For the cancellation to occur, it is essential to choose the two EOW branes to have the same tension, ensuring that the resulting boundary entropies are identical. As we have shown that we can decompose the banana into sections of circles which  slices of hemispheres. It is more transparent to consider the cone geometry where on each slice of constant $\phi$, we have a wedge configuration. Therefore, the cancellation of the surface integrals in banana geometry happens exactly similar to \eqref{bridgec}. We also checked this cancellation perturbatively by considering a small $b_x$ parameter. 

\end{document}